\documentclass[14 pt]{article}
\usepackage[utf8]{inputenc}
\usepackage[backend=biber,style=numeric-comp,sorting=none]{biblatex}
\usepackage{amsmath}
\usepackage{amsthm}
\usepackage{amssymb}
\usepackage{graphicx}
\usepackage{slashed}
\usepackage{MnSymbol}
\usepackage{mathtools}
\usepackage{cancel}
\usepackage{xcolor}
\usepackage{simpler-wick}
\usepackage{hyperref} 
\hypersetup{backref=true,       
    pagebackref=true,               
    hyperindex=true,                
    colorlinks=true,                
    breaklinks=true,                
    urlcolor= black,                
    linkcolor= blue,                
    bookmarks=true,                 
    bookmarksopen=false,
    filecolor=black,
    citecolor=blue,
    linkbordercolor=blue
}
\numberwithin{equation}{section}
\newtheorem{theorem}{Theorem}
\theoremstyle{definition}

\newcommand{\efill}{\;\;\;\;\;\;\;\;\;\;}
\newcommand{\be}{\begin{equation}}

\newcommand{\ee}{\end{equation}}
\newcommand{\bthe}{\begin{theorem}}
\newcommand{\ethe}{\end{theorem}}
\newcommand{\bea}{\begin{eqnarray}}
\newcommand{\eea}{\end{eqnarray}}
\newcommand{\al}{\alpha}
\renewcommand{\d}{\delta}
\newcommand{\e}{\epsilon}
\newcommand{\G}{\Gamma}

\renewcommand{\k}{\kappa}
\newcommand{\La}{\Lambda}
\newcommand{\la}{\lambda}
\newcommand{\m}{\mu}
\newcommand{\n}{\nu}
\newcommand{\Om}{\Omega}

\newcommand{\s}{\sigma}

\newcommand{\hlf}{\frac{1}{2}}
\newcommand{\F}{\mathcal{F}}

\newcommand{\non}{\nonumber}

\newcommand{\p}{\partial}

\newcommand{\R}{\mathbb{R}}
\newcommand{\rr}{\rightarrow}
\newcommand{\w}{\wedge}
\newcommand{\Z}{\mathbb{Z}}

\renewcommand{\O}{\mathcal{O}}

\newcommand{\lp}{\left(}
\newcommand{\rp}{\right)}
\newcommand{\ls}{\left[}
\newcommand{\rs}{\right]}

\title{$O(16)\times O(16)$ heterotic theory on $AdS_3\times S^3\times T^4$}
\author{Hassaan Saleem}
\begin{document}
\begin{titlepage}

\begin{center}
\hfill         \phantom{xxx}  

\vskip 2 cm {\Large \bf $O(16)\times O(16)$ heterotic theory on $AdS_{3}\times S^{3}\times T^{4}$} 
\vskip 1.25 cm {\bf Daniel Robbins and Hassaan Saleem\non\\
\vskip 0.2 cm
 
{\it Department of Physics, University at Albany, Albany, NY, USA}

\vskip 0.2 cm
}
\end{center}
\vskip 1.5 cm
\begin{abstract}
In this paper, we study non-supersymmetric $AdS_{3}\times S^{3}$ vacuua of $O(16)\times O(16)$ heterotic theory on a string scale $T^{4}$ background, which are parameterized by a pair of flux integers.  Adding the one-loop scalar potential to the effective theory contributes positively to the cosmological constant, but we find that there is no uplift to de Sitter for any values of the fluxes.  We study the fluctuations around these vacua and show that all scalar and tensor modes from the six-dimensional effective theory lie above the Breitenlohner-Freedman bound.  The moduli coming from the torus compactification will also be above the bound, at least for a large range of fluxes.
\baselineskip=18pt
\end{abstract}
\end{titlepage}
\tableofcontents
\newpage

\section{Introduction}

We have known for the last two and a half decades that the universe is expanding with an accelerated expansion \cite{Planck:2018vyg}. One way to explain this observation is to assume that our universe has a positive cosmological constant, making it a de Sitter universe. Since string theory is a proposed framework for fundamental interactions, one expects to have some string construction that gives stable (or at least long lived) de Sitter vacuua. After decades of studies (see~\cite{Silverstein:2001xn, Maloney:2002rr, Kachru:2003aw, Flauger:2008ad, Rummel:2011cd, Achucarro:2015kja, Gallego:2017dvd, Kachru:2019dvo, Bernardo:2020lar, Demirtas:2021nlu, Shukla:2022srx, McAllister:2024lnt} for a sample) we have learned that this is not an easy task~\cite{Dine:2020vmr, Bena:2023sks}. 

Some have questioned if string theory even has de Sitter vacua~\cite{Danielsson:2018ztv, Obied:2018sgi, Ooguri:2018wrx}. In addition, there are some results that prohibit a de Sitter construction in certain scenarios~\cite{Maldacena:2000mw, Flauger:2008ad, Kutasov:2015eba} or question the validity of existing proposals~\cite{Lust:2022lfc} . For most of the attempts that we have mentioned above, supersymmetry is found at the string scale, but one has to break supersymmetry at low energies (e.g. by fluxes~\cite{Sethi:2017phn}) if one needs a phenomenologically relevant model. Instead, it is interesting to ask about models where (spacetime) supersymmetry is broken already at the string scale.

Of course string theories without spacetime supersymmetry are under less control than supersymmetric string theories.  Although we will maintain supersymmetry on the world-sheet, the lack of supersymmetry in spacetime introduces a number of potential issues that need to be approached cautiously.

In supersymmetric backgrounds, the dilaton $\phi$ will often remain a flat direction (i.e.~with no potential) but without supersymmetry, quantum corrections can and typically will generate a potential for $\phi$. If it's not stabilized, the dilaton can run away to strong coupling (breaking perturbation theory) or zero coupling (making the model unrealistic).  Additionally, non-supersymmetric string theories often suffer from tachyons in their spectrum, and hence perturbative instabilities, which are not present in supersymmetric models.

Finally, there are conjectures that non-supersymmetric $AdS$ is unstable (the instability may be perturbative or non-perturbative)~\cite{Ooguri:2016pdq, Freivogel:2016qwc} but there are some counterexamples which may challenge this claim~\cite{Guarino:2020flh, Giambrone:2021wsm, Eloy:2021fhc}.

Despite the fact that it is hard to get rid of tachyons in non-supersymmetric models, people have come up with perturbatively stable string models where supersymmetry is broken at the string scale. These models include the closed string $O(16)\times O(16)$ heterotic string theory~\cite{Dixon:1986iz, Ginsparg:1986wr, Alvarez-Gaume:1986ghj} and two orientifold models i.e.~the $USp(32)$ Sugimoto model~\cite{Sugimoto:1999tx} and the $U(32)$ model~\cite{Sagnotti:1995ga, Sagnotti:1996qj}. All of these models can be constructed by starting with one of the tachyon-free supersymmetric string theories and then breaking supersymmetry with some orbifold/orientifold. For a review of these models, see~\cite{Basile:2021vxh, Mourad:2017rrl, Leone:2025mwo, Dudas:2025ubq}.  These models have been used to construct interesting backgrounds, which include perturbatively stable vacuua with exponentially suppressed cosmological constant \cite{Abel:2015oxa}, smooth and orientifold Calabi-Yau compactifications with Standard Model-like models \cite{Blaszczyk:2015zta, Blaszczyk:2014qoa}, $AdS_{7}\times S^{3}$ Freund-Rubin compactifications \cite{Mourad:2016xbk, Basile:2018irz}, warped and time dependent solutions \cite{Mourad:2021roa}, brane-like and domain wall solutions \cite{Mourad:2024dur}, and backgrounds of the form $AdS_{4}\times X_{3}\times Y_{3}$ and $AdS_{5}\times T_{p,q}$\cite{Raucci:2025bev}.

When working with one of these string theories, one can consider the one loop correction to the potential in the effective field theory (which was absent in supersymmetric string theories) and hope that if one starts with an $AdS$ spacetime, this one-loop contribution may uplift the cosmological constant from a negative value to a positive value, giving us a tachyon free de Sitter construction. Moreover, one has to check whether the scalars present in the effective gravity model go below the Breitenlohner-Freedman (BF) bound when the one-loop effects are included. There have been significant developments on these questions. \cite{Basile:2020mpt} proved a no-go theorem along the lines of \cite{Maldacena:2000mw} to show that for a fairly large class of compactifications, deSitter and Minkowski solutions are ruled out. This no-go theorem assumes an absence of electric fluxes and therefore, ~\cite{Baykara:2022cwj} performed an analysis for $AdS_{3}\times S^{3}\times S^{3}\times S^{1}$ solution with electric fluxes, making use of the analysis performed in~\cite{Eberhardt:2017fsi}. It was found that neither the uplift to deSitter nor a violation of the BF bound takes place. Later,~\cite{Fraiman:2023cpa} found eight tachyon-free maximal symmetry enhancement points in the circle moduli space, but they were all unstable.

In this paper, we analyze the $AdS_{3}\times S^{3}$ background of the $O(16)\times O(16)$ theory on a string scale $T^{4}$ and study the spectrum around a critical point in the $T^{4}$ moduli space. In our construction, there are two $H_{3}$ fluxes $(n_{1},n_{5})$ on $AdS_{3}$ and $S^{3}$ respectively. At tree-level, the string coupling is frozen at the following value
\be
g^{2}_{s}=\frac{v|n_{5}|}{(2\pi)^{4}|n_{1}|}
\ee
where $v$ is the volume of $T^{4}$ in string units. In the $n_{1}\rr 0$ limit, $g_{s}$ blows up. However, as can be seen in~\eqref{small n1 expansions}, $g_{s}$ is stabilized at a finite value for $n_{1}=0$ when we add the one-loop correction, which is the same behavior as seen in~\cite{Baykara:2022cwj}.  Therefore, for $n_{1}=0$, we have intrinsically quantum vacua for different values of $n_{5}$, but we will find that strictly in this limit, we run into trouble with scalars from the $T^4$ moduli dropping below the BF bound.   Addressing the question of whether the one-loop potential can uplift the tree-level vacuum to de Sitter, we conclude that this uplift never happens for the $T^{4}$ compactification, and we also provide a slightly generalized argument along the lines of \cite{Basile:2020mpt} to show that the absence of this uplift shouldn't be surprising.  The analysis of fluctuations around the vacuum solution reveals that there are no tachyons below the BF bound or other perturbative instabilities coming from the six-dimensional supergravity fields, and, at least for a range of fluxes, there are no tachyons below the BF bound coming from the torus moduli either.

This paper is structured as follows: In section 2, we derive the tree level potential and find its extremum to find the values of $g_{s}$ and the $AdS_{3}$ cosmological constant $\Lambda_{3}$. In section 3, we include the one-loop correction to the potential (assuming that the size of $AdS_{3}\times S_{3}$ is much bigger than $T^{4}$) and find the one-loop corrected results for $g_{s}$ and $\Lambda_{3}$. We also explore different limiting cases in this section. In section $4$, we perform the perturbative stability analysis and find that there are no scalars that violate the BF bound.  In the appendices, we present some lattice sum calculations to show that a square torus at self-dual radius is an extremum of the scalar potential, and to compute the magnitude of the one-loop contribution for this specific case.

\section{Tree level analysis}\label{sec:TreeLevel}

Adopting the conventions of \cite{Baykara:2022cwj}, the tree level action of $O(16)\times O(16)$ non supersymmetric heterotic theory is
\be\label{heterotic theory action}
S=\frac{1}{2\k^{2}_{10}}\int d^{10}x\;e^{-2\phi}\sqrt{-g}\lp R+4\lp\partial \phi\rp^{2}-\frac{1}{12}|H_{3}|^{2}-\frac{1}{4}|F_{2}|^{2}\rp,
\ee
with
$$
\k^{2}_{10}=\hlf (2\pi)^{7}\al'^{4}e^{2\phi_{0}},
$$
where $\phi_{0}$ is the vev of the dilaton, which means that $\phi$ in the action is a fluctuation with zero vacuum expectation value i.e. $\langle \phi\rangle=0$. Compactifying on a $T^{4}$, the volume integral measure becomes
\be
\frac{1}{2\k^{2}_{10}}\int d^{10}x\;e^{-2\phi}\sqrt{-g}=\frac{\text{vol}(T^{4})}{2\k^{2}_{10}}\int d^{6}x\;e^{-2\phi}\sqrt{-g_{6}}
\ee
where $\text{vol}(T^{4})$ is the volume of $T^{4}$. Now, we want to study $AdS_{3}\times S^{3}$ vacuua of this compactified theory, and we start by quoting the line element for $AdS_{3}\times S^{3}$, which is
\be
ds^{2}_{6}=ds^{2}_{AdS_{3}}+e^{2\chi}d\Om^{2}_{3},
\ee
where $e^{\chi}$ is the radius modulus of $S^{3}$. $d\Om_{3}^{2}$ is the metric on $S^{3}$ with radius $L$ (and volume $2\pi^{2}L^{3}$). Compactifying on $S^{3}$, the volume integral measure becomes
\be
\frac{1}{2\k^{2}_{3}}\int d^{3}x\;\mathcal{V}\sqrt{-g_{3}},
\ee
where $g_{3}$ is the determinant of the metric on $AdS_{3}$ and
\be\label{Einstein frame def}
\mathcal{V}=e^{-2\phi+3\chi},\efill \k_{3}^{2}=\frac{\k^{2}_{10}}{2\pi^{2}L^{3}\times \text{vol}(T^{4})}.
\ee
If we go to the Einstein frame using $g_{3}=\mathcal{V}^{-2}\hat{g}_{3}$, we get the following two terms from the Ricci scalar term in \eqref{heterotic theory action}
\be
\int d^{3}x\;\mathcal{V}\sqrt{-g_{3}} R_{3}=\int d^{3}x\;\sqrt{-\hat{g}_{3}}\lp \hat{R}_{3}+\mathcal{V}^{-2} R_{7}\rp.
\ee
The second term will be required to construct the tree-level potential. $R_{7}$ is given in terms of the radius of the sphere as follows
\be\label{Ricci scalar sphere}
R_{7}=R_{S_{3}}+R_{T_{4}}=\frac{6}{L^{2}}e^{-2\chi}+0=\frac{6}{L^{2}}e^{-2\chi}.
\ee
Now, let's attend to the $H_{3}$ field. The magnetic flux $H_3^m$ is the $H_3$ flux wrapping the $S^3$ which is quantized as
\be\label{magnetic H3}
\frac{1}{(2\pi)^{2}\al'}\int_{S^{3}}H^{m}_{3}=n_{5}
\quad\Rightarrow\quad H_{3}^{m}=\frac{2\al'n_{5}}{L^{3}}\e_{S^{3}}\quad\Rightarrow\quad |H^{m}_{3}|^{2}=\frac{24\al'^{2}n^{2}_{5}}{L^{6}}e^{-6\chi},
\ee
where $n_{5}\in \Z$ and $\e_{S^3}$ is the volume form for the $S^3$ of radius $L$. To determine the electric $H_{3}$ flux (i.e.~$H^{e}_{3}$), we determine the Poincar\'e dual flux $H_{7}$ first. 
The correct Poincar\'e dual in this case is $H_7=\ast(e^{-2\phi}H_3)$ (since this is the 7-form that is closed by the equation of motion), and its quantization condition is
\be
\frac{1}{(2\pi)^{6}\al'^{3}g^{2}_{s}}\int_{S^{3}\times T^{4}}H_{7}=n_{1}\quad\Rightarrow\quad H_{7}=\frac{(2\pi)^{6}\al'^{3}n_{1}g^{2}_{s}}{2\pi^{2}L^{3}\times \text{vol}(T^{4})}\e_{S^{3}}\w\e_{T^{4}}=\frac{32\pi^{4}\al'n_{1}g^{2}_{s}}{vL^{3}}\e_{S^{3}}\w\e_{T^{4}},
\ee
where $\text{vol}(T^{4})$ is written as $v\al'^{2}$ with $v$ a dimensionless parameter. Moreover, $\e_{T^4}$ is the volume form for $T^4$. Transforming back to $H_3^e$ we get
\be\label{electric H3}
H^{e}_{3}=\mathcal{V}^{-1}\frac{32\pi^{4}\al'g^{2}_{s}n_{1}}{vL^{3}}\e_{3}\quad\Rightarrow\quad |H^{e}_{3}|^{2}=\mathcal{V}^{-2}\frac{6144\pi^{8}\al'^{2}g^{4}_{s}n^{2}_{1}}{v^{2}L^{6}}.
\ee
Now, the potential term in the Einstein frame comes from the following three terms
\be
-\frac{1}{2\k^{2}_{3}}\int d^{3}x\sqrt{-\hat{g}_{3}}V_{\text{tree}}(\phi,\chi)=\frac{1}{2\k^{2}_{3}}\int d^{3}x\;\mathcal{V}^{-2}\sqrt{-\hat{g}_{3}}\lp R_{7}-\frac{1}{12}|H^{m}_{3}|^{2}-\frac{1}{12}|H^{e}_{3}|^{2}\rp,
\ee
resulting in
\be
\label{eq:Vtree}
V_{\text{tree}}(\phi,\chi)=\mathcal{V}^{-2}\ls -\frac{6}{L^{2}}e^{-2\chi}+\frac{2\al'^{2}n^{2}_{5}}{L^{6}}e^{-6\chi}+\mathcal{V}^{-2}\frac{512\pi^{8}\al'^{2}g^{4}_{s}n^{2}_{1}}{v^{2}L^{6}}\rs.
\ee
The partial derivatives of this potential are given as
\begin{align}\label{tree phi der}
\p_{\phi}V_{\text{tree}}=\ & -\frac{8e^{4(-3\chi+\phi)}}{L^{6}v^{2}}\lp 3L^{4}v^{2}e^{4\chi}-512g^{4}_{s}n^{2}_{1}\pi^{8}\al'^{2}e^{4\phi}- n^{2}_{5}v^{2}\al'^{2}\rp,\\
\label{tree chi der}
\p_{\chi}V_{\text{tree}}=\ & \frac{24e^{4(-3\chi+\phi)}}{L^{6}v^{2}}\lp 2L^{4}v^{2}e^{4\chi}-256g^{4}_{s}n^{2}_{1}\pi^{8}\al'^{2}e^{4\phi}-n^{2}_{5}v^{2}\al'^{2}\rp.
\end{align}
Note that we haven't set any of the moduli to zero yet. Setting the derivatives in \eqref{tree phi der} and \eqref{tree chi der} to zero with vanishing fluctuating moduli (i.e. $\phi=\chi=0$) results in
\be\label{tree L sol}
L^{4}=n^{2}_{5}\al'^{2},
\ee
\be\label{tree gs sol}
g_{s}^{2}=\frac{v |n_{5}|}{(2\pi)^{4}|n_{1}|}.
\ee
The value of the potential for $\phi=\chi=0$, at the extremum is
\be
V_{\text{min}}=-\frac{2}{\al'|n_{5}|},
\ee
which in turn corresponds to a cosmological constant
\be
\Lambda_3=\hlf V_{\text{min}}=-\frac{1}{\al'|n_{5}|}.
\ee
This $\Lambda_3$ is negative.

The value of $v$ depends on the volume of the $T^{4}$ on which we compactify. A particular case is a square torus at the self-dual radius, in which case $v=(2\pi)^4$ and we have;
\be
g^{2}_{s}=\left|\frac{n_{5}}{n_{1}}\right|.
\ee

\section{One loop correction}

\subsection{One loop corrected potential}

To include the one-loop effects, we need to find the one-loop correction to the potential, given by integrating the genus one partition function of the $O(16)\times O(16)$ heterotic string compactified on $T^4$ over the fundamental domain of the genus one moduli space. We will assume that the fluxes are such that the AdS length scale and the radius of $S^{3}$ are much bigger than the self-dual radius, which enables us to approximate the $AdS_{3}\times S^{3}$ background by a $\R^{1,5}$ background. The six-dimensional cosmological constant for toroidal compactifications is given as follows (following~\cite{Fraiman:2023cpa} with $d=4$)
\begin{equation}
\La_{6}=-\frac{1}{2\lp 2\pi\sqrt{\al'}\rp^{6}}\int _{\mathcal{F}}\frac{d^{2}\tau}{\tau^{2}_{2}}Z(\tau).
\end{equation}

Alternatively we can interpret this as a contribution to the six-dimensional scalar potential,
\begin{equation}\label{six dim pot def}
    -\int d^{6}x \sqrt{-g}\Lambda_{6}
    =-\frac{1}{2\k^{2}_{6}}\int d^{6}x  \sqrt{-g}\;V^{(6)},
\end{equation}
where
\begin{equation}
    V^{(6)}=2\k^{2}_{6}\Lambda_{6},\efill\efill\k^{2}_{6}=\frac{\k^{2}_{10}}{\text{vol}(T^{4})}=\frac{(2\pi)^{7}\al'^{2}}{2v}g^{2}_{s}.
\end{equation}
When we compactify~\eqref{six dim pot def} on $S^{3}$, we get
\begin{equation}
    -\frac{1}{2\k^{2}_{6}}\int d^{6}x\;\;\sqrt{-g}\;\;V^{(6)}=-\frac{1}{2\k^{2}_{6}}(2\pi^{2}L^{3})\int d^{3}x\;\;\sqrt{-g}\;\;e^{3\chi} V^{(6)}=-\frac{1}{2\k^{2}_{3}}\int d^{3}x\;\;\sqrt{-g}\;e^{3\chi}\;V^{(6)}.
\end{equation}
Going to the Einstein frame, this term becomes
\begin{equation}
    -\frac{1}{2\k^{2}_{3}}\int d^{3}x\;\;\sqrt{-\hat{g}}\;\mathcal{V}^{-3}e^{3\chi}\;V^{(6)},
\end{equation}
resulting in the following one-loop potential in three dimensions
\be\label{one loop potential}
V_{1-\text{loop}}=\mathcal{V}^{-3}e^{3\chi}V^{(6)}=2\underbrace{\lp\frac{(2\pi)^{7}(\alpha')^3}{2v}\Lambda_6\rp}_{\la}\mathcal{V}^{-3}e^{3\chi}\frac{g^{2}_{s}}{\al'}=2\la\mathcal{V}^{-3}e^{3\chi}\frac{g^{2}_{s}}{\al'},
\ee
where the dimensionless quantity $\la$ is defined as
\be\label{lambda general expression}
\la=\frac{(2\pi)^{7}(\alpha')^3}{2v}\Lambda_6=-\frac{\pi}{2v}\int_{\F} \frac{d^{2}\tau}{\tau^{2}_{2}}Z(\tau).
\ee
On general grounds we expect $\la$ to be an order one number.  Indeed, the value of $\lambda$ for the square torus is calculated in Appendix~\ref{Appendix C}, yielding $\la\simeq 1.321$.

The one-loop contribution to the potential is given in~\eqref{one loop potential}. The new potential is $V=V_{\text{tree}}+V_{1-\text{loop}}$. For $\phi=\chi=0$, the $\p_{\phi}V$ and $\p_{\chi}V$ derivatives being zero give the following equations
\begin{align}
\label{one loop phi der}
\p_{\phi}V=0\quad\Rightarrow\quad & 2\al'^{2}\lp n^{2}_{5}+\frac{2(2\pi)^{8}g^{4}_{o}n^{2}_{1}}{v^{2}}\rp-6L^{4}_{o}+\frac{3(2\pi)^{4}\la g^{2}_{o}L^{6}_{o}}{\al'v}=0,\\
\label{one loop chi der}
\p_{\chi}V=0\quad\Rightarrow\quad & 2\al'^{2}\lp n^{2}_{5}+\frac{(2\pi)^{8}g^{4}_{o}n^{2}_{1}}{v^{2}}\rp-4L^{4}_{o}+\frac{(2\pi)^{4}\la g^{2}_{o}L^{6}_{o}}{\al'v}=0.
\end{align}
where $L_{o}$ and $g_{o}$ are the radius of the sphere and string coupling that solve~\eqref{one loop phi der}, and~\eqref{one loop chi der}. To simplify, we will set $v=(2\pi)^4$, its value for the square torus at self-dual radius.  The $v$ dependence can easily be restored later by rescaling $g_o^2$, since $v$ only appears in the combination $(2\pi)^4g_o^2/v$.

If we use~\eqref{one loop phi der} and~\eqref{one loop chi der} to get rid of the $g_{o}^{2}L^{6}_{o}$ term, we get
\be\label{go^2 sol} 
L^{4}_{o}=\lp\frac{n^{2}_{1}g^{4}_{o}}{3}+\frac{2n^{2}_{5}}{3}\rp\al'^{2}\Rightarrow g^{2}_{o}= \frac{1}{|n_{1}|\al'}\sqrt{3L^{4}_{o}-2\al'^{2}n^{2}_{5}}.
\ee
Therefore, we now have to solve~\eqref{one loop phi der} and~\eqref{go^2 sol} for $L^{2}_{o}$ and $g^{2}_{o}$. 
Moving the $g_o^2L_o^6$ term in~\eqref{one loop phi der} to one side, then squaring both sides of the equation, and then using~\eqref{go^2 sol} to substitute $g_o^4$ in terms of $L_o^4$ yields a quartic equation for $L_o^4$, which can thus be solved exactly.

The physically relevant branch solution for $L^{2}_{o}$ is given by
\be\label{Lo^2 sol}
L^{2}_{o}=\begin{cases}\frac{1}{3 \sqrt{2}}\ls 3 \al'^{2}n^{2}_{5}+\sqrt{3(\mu_{1}+\mu_{2})}+\sqrt{6\mu_{1}-3\mu_{2}+\frac{18 \sqrt{3}\;n^{2}_{5}\al'^{6} }{\sqrt{\mu_{1}+\mu_{2}}}\left(n^{4}_{5}-\frac{60 n^{2}_{1}}{\la^{2}}\right)}\rs^{1/2}, & \frac{\la n^{2}_{5}}{n_{1}}\geq 60, \\
\frac{1}{3 \sqrt{2}}\ls 3 \al'^{2}n^{2}_{5}-\sqrt{3(\mu_{1}+\mu_{2})}+\sqrt{6\mu_{1}-3\mu_{2}-\frac{18 \sqrt{3}\;n^{2}_{5}\al'^{6} }{\sqrt{\mu_{1}+\mu_{2}}}\left(n^{4}_{5}-\frac{60 n^{2}_{1}}{\la^{2}}\right)}\rs^{1/2}, & \frac{\la n^{2}_{5}}{n_{1}}\leq 60,
\end{cases}
\ee
where
\be
\mu_{1}=3\al'^{4}n^{4}_{5}+\frac{24n^{2}_{1}\al'^{4}}{\la^{2}},\efill\mu_{2}=\zeta\sqrt[3]{4}+\frac{72\sqrt[3]{2}n^{2}_{1}\al'^{8}}{\la^{4}\zeta}\lp n^{2}_{1}-6n^{4}_{5}\la^{2}\rp,
\ee
\be
\zeta=\frac{\al'^{4}}{\la^{2}}\lp -1458n^{8}_{5}n^{2}_{1}\la^{4}+3888n^{4}_{5}n^{4}_{1}\la^{2}-432n^{6}_{1}+162n^{2}_{1}n^{4}_{5}\la^{2}\sqrt{81n^{8}_{5}\la^{4}-144n^{4}_{1}+1104n^{2}_{1}n^{4}_{5}\la^{2}}\rp^{1/3},
\ee
with the solution for $g^{2}_{o}$ then given by~\eqref{go^2 sol}. 

These results are plotted in Figure~\ref{fig:Lo, go and Vmin plots}, and show that the quantities are well-behaved for the whole range of available fluxes. Figure~\ref{fig:Lo, go and Vmin plots} also shows that the one-loop correction can't uplift $V_{\text{min}}$ to positive values, which means that we don't get a deSitter uplift. The results in the $n_{1}\rr 0$ limit are consistent with the no-go theorem in \cite{Basile:2020mpt}. Since the assumptions of the no-go theorem in \cite{Basile:2020mpt} include an absence of electric flux, this theorem doesn't automatically rule out deSitter for arbitrary $n_{1}$. However, we argue below \eqref{eq:BEOM} that the procedure to prove this theorem can be repeated to show that even for the case that we are studying (which includes an electric flux), one can't get deSitter. Therefore, the lack of a deSitter uplift isn't surprising.
\begin{figure}
    \centering
    \includegraphics[width=50 mm]{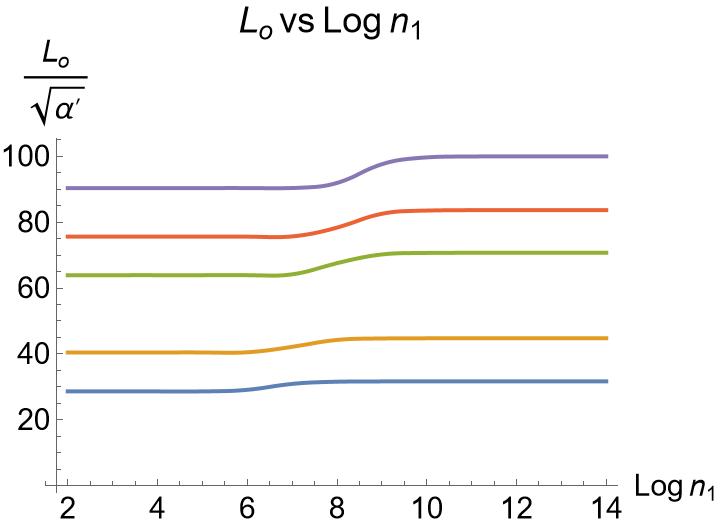}\hfill
    \includegraphics[width=50 mm]{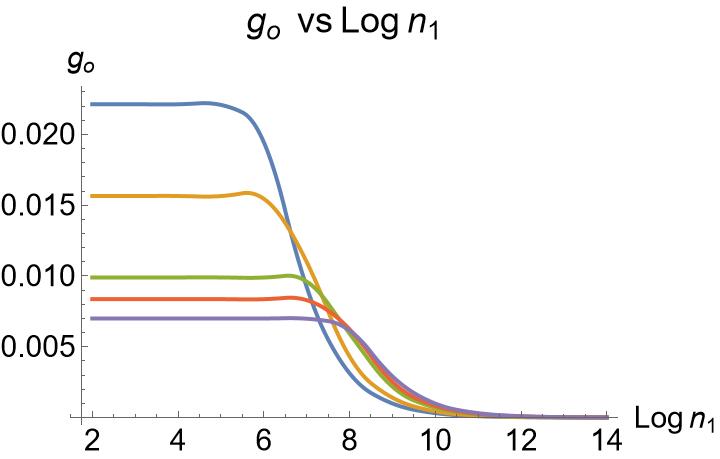}\hfill
    \includegraphics[width=50 mm]{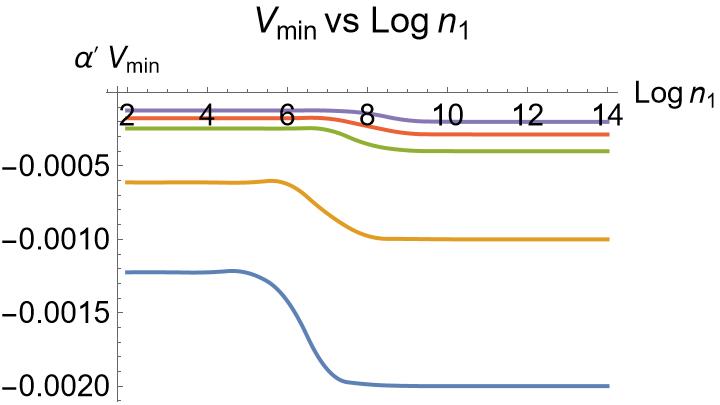}\hfill
    \includegraphics[width=12mm]{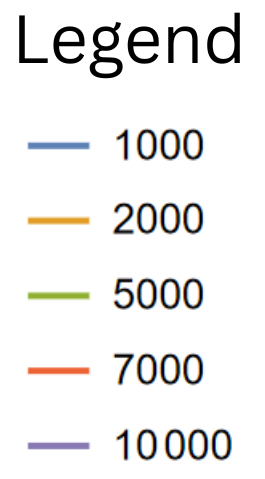}
    \caption{Plots of $L_{o}, g_{o}$ and $V_{\text{min}}$ against $\log n_{1}$ for different values of $n_{5}$ given in the legend.}
    \label{fig:Lo, go and Vmin plots}
\end{figure}

Note that if we look at the ratios of one-loop to tree-level quantities, $\tilde{g}=g_o/g_s$ and $\tilde{L}=L_o/L$, then these only depend on the parameters $\lambda$, $n_1$, and $n_5$ through the combination $s:=\lambda n_5^2/|n_1|$.  The limit $s\rightarrow 0$, corresponding to sending $\lambda\rightarrow 0$ and hence turning off the one-loop potential (but which can also be viewed as either a large $|n_1|$ or small $|n_5|$ limit, holding other quantities fixed), recovers the tree-level results.

We can expand~\eqref{Lo^2 sol},~\eqref{go^2 sol} and~\eqref{LoAdS^2 expression} as corrections to the tree-level solution, $L^2=\al'|n_5|$, $g_s^2=(v/(2\pi)^4)|n_5/n_1|$, $\La_3=-1/(\al'|n_5|)$.  The small $s$ expansions can be written as

\begin{align}
\label{Lo tree expansion}
L^{2}_{o}=\ & L^{2}\lp 1-\frac{1}{4}s+\frac{11}{32}s^2-\frac{73}{128}s^3+\cdots\rp,\\
\label{go tree expansion}
g^{2}_{o}=\ & g^{2}_{s}\lp 1-\frac{3}{4}s+\frac{27}{32}s^2-\frac{171}{128}s^3+\cdots\rp,\\
\label{Vmin tree expansion}
\La_{3,o}=\ & \La_3\lp 1-\frac{1}{4}s+\frac{3}{32}s^2-\frac{1}{128}s^3+\cdots\rp.
\end{align}

Note that the expansion parameter can also be expressed in terms of the tree-level quantities and $\la$ as
\be
s=\frac{\la n_5^2}{|n_1|}=\frac{\lp 2\pi\rp^4}{v}\frac{g_s^2L^2\la}{\al'}.
\ee
Of course, trusting the corrections beyond first order in $s$ may seem suspect since at that point we should presumably include two-loop corrections to the effective action.  In fact, we will argue below in section~\ref{subsec:ScalingArguments} that these corrections can be meaningful as long as $n_5$ is large.  It is also worth noting that the $-1/4$ appearing in the first order correction in $\La_{3,o}$ is identical (for $v=(2\pi)^4$) to the result found in~\cite{Baykara:2022cwj}.

\subsection{Large $s$ limit}
\label{subsec:LargesLimit}

The other limit, $s\rightarrow\infty$ is also of interest.  We can think of this as a limit in which we turn off the electric flux, sending $n_1\rightarrow 0$.  At tree-level, this limit is badly behaved, with the string coupling diverging and no solution available in the strict $n_1=0$ limit.  Once we include the one-loop contribution, however, the situation improves; the divergence in $g_o$ gets cut off, and the quantities $g_o$, $L_o$, and $V_{\text{min}}$ all asymptote to constant non-zero values in this limit (corresponding to the left-hand side of the plots in Figure~\ref{fig:Lo, go and Vmin plots}).  In~\cite{Baykara:2022cwj} these were referred to as {\it{intrinsically quantum}} solutions.

The large $s$ expansions of \eqref{Lo^2 sol} and \eqref{go^2 sol} are as follows
\be\label{small n1 expansions}
L^{2}_{o}=\sqrt{\frac{2}{3}}\al' |n_{5}|\lp 1+\frac{3}{8}s^{-2}-\frac{261}{128}s^{-4}+...\rp,\efill g^{2}_{o}=\sqrt{\frac{3}{2}}\frac{1}{|\la n_{5}|}\lp 1-\frac{21}{8}s^{-2}+ \frac{2115}{128}s^{-4}+...\rp.
\ee

The corresponding cosmological constant in the $n_1\rr 0$ limit is
\be
\label{LoAdS^2 expression}
\lim_{n_1\rr 0}\Lambda_{3,o}=\lim_{n_1\rr 0}\lp -\frac{1}{L_{AdS,o}^2}\rp=\lim_{n_1\rr 0}\lp -L_o^{-2}+\frac{\la g_o^2}{2\al'}\rp=-\hlf\sqrt{\frac{3}{2}}\frac{1}{|n_5|\al'}<0,
\ee
which is negative. Therefore, we do not find an uplift to de Sitter.

\subsection{Scaling arguments and higher corrections}
\label{subsec:ScalingArguments}

Let's take a moment to briefly present some scaling arguments and comments about corrections that we have neglected.  At tree-level, we find curvatures of AdS$_3$ and $S^3$ that go like Riemann~$\sim 1/(\alpha'|n_5|)$, and a string coupling that scales as $g_s\sim|n_5/n_1|^{1/2}$.  These are obtained by minimizing the effective potential $V_{\text{tree}}$, \eqref{eq:Vtree}, and at the minimum the various terms in $V_{\text{tree}}$, due to curvature, magnetic flux, and electric flux, are all of the same order, $1/(\alpha'|n_5|)$.

Now we add in the one-loop correction to the effective potential, which goes like (recall that $\lambda$ is an order one dimensionless number, so we suppress it) $g_s^2/\alpha'$.  If we assume that it doesn't shift the minimum much, then this scales like $s/(\alpha'|n_5|)$, where $s\sim n_5^2/|n_1|$.  As long as $s$ is small, then the assumption that this won't shift the minimum much will be valid.  So if we take flux numbers satisfying $|n_1|\gg n_5^2\gg 1$, then both $s$ and the curvatures (in string units) will be small.  In this regime, the terms we have neglected include both higher-derivative corrections such as $\alpha'(\text{Riemann})^2$, as well as higher loop contributions that go like $g_s^4/\alpha'$, the leading genus-two contribution to the effective potential, and also terms that are higher order in both the $\alpha'$ and $g_s$ expansions.  These neglected terms scale as
\begin{align}
    \alpha'\lp\text{Riemann}\rp^2\sim\ & \frac{1}{\alpha'n_5^2},\\
    \frac{g_s^4}{\alpha'}\sim\ & \frac{s^2}{\alpha'n_5^2}.
\end{align}
So if we indeed have $|n_1|\gg n_5^2\gg 1$, then these terms are all suppressed by at least an extra factor of $1/|n_5|$.

But we claim that the situation is actually even better than this; we don't actually need $s$ to be small.  In this case the one-loop correction to the potential does shift the location of the minimum by a significant amount, but the shift remains finite.  Even in the $s\rightarrow\infty$ limit discussed in section~\ref{subsec:LargesLimit}, we find curvatures that still scale as Riemann~$\sim 1/(\alpha'|n_5|)$ and a string coupling that goes as $g_o\sim|n_5|^{-1/2}$.  As $s$ interpolates between $0$ and $\infty$, these values change monotonically.  Even in this large $s$ regime the neglected higher order terms scale as
\begin{align}
    \alpha'\lp\text{Riemann}\rp^2\sim\ & \frac{1}{\alpha'n_5^2},\\
    \frac{g_s^4}{\alpha'}\sim\ & \frac{1}{\alpha'n_5^2}.
\end{align}
Comparing to the terms in $V_{tree}$, which for large $s$ will scale as either $1/(\alpha'|n_5|)$ (for curvature and magnetic flux contributions) or $1/(\alpha'|n_5|s^2)$ (for the electric flux contribution), we see that, as long as $|n_5|$ is large, regardless of the value of $s$, it is reasonable to drop the higher order terms.

\section{Stability analysis}
\subsection{Vacuum solution}\label{subsec: vacuum sol}

In this section, we will study fluctuations around the previously determined background solution in the effective six-dimensional gravitational theory.  Our analysis follows earlier work, specifically~\cite{Deger:1998nm,Eberhardt:2017fsi,Baykara:2022cwj}.  The effective action in six dimensions takes the form
\begin{equation}\label{6D string frame action}
    S_6=\frac{1}{2\kappa_6^2}\int d^6x\sqrt{-g}\left[e^{-2\phi}\left(R+4\left(\partial\phi\right)^2-\frac{1}{12}\left|H_3\right|^2-\frac{1}{4}\left|F_2\right|^2-\hlf\left|\partial\sigma\right|^2\right)-\frac{2\lambda g_s^2}{\alpha'}-\hlf\mu_{\alpha\beta}\sigma^\alpha\sigma^\beta\right].
\end{equation}
Here $g_{MN}$ is the six-dimensional metric, $\phi$ is the dilaton, $H_{MNP}$ is the three-form field strength in six dimensions, $F^i_{MN}$ are the field strengths for the twenty-four vector fields (sixteen from the gauge fields of the ten-dimensional theory, and eight more from reducing the metric and $B$-field on $T^4$), and $\sigma^\alpha$ are the eighty scalars coming from the Narain moduli of the $(20,4)$ signature lattice.  The indices $M$ run over the six dimensions of space-time.  The one-loop contribution includes the six-dimensional cosmological constant piece $\Lambda:=2\lambda g_s^2/\alpha'$, as well as the mass terms for the $\sigma^\alpha$ (so in particular all of the eigenvalues of $\mu_{\alpha\beta}$ will have an explicit factor of $g_s^2$).

Note that we normalize the string coupling $g_s=g_\circ$ such that the field $\phi$ has zero vacuum expectation value.  Similarly, the fields $\sigma^\alpha$ will have zero VEV, and we only keep the terms in the action quadratic in these fields.  They can then be chosen so that their kinetic terms are canonically normalized and their mass matrix $\mu_{\alpha\beta}$ is diagonal.  We call these, along with the three-dimensional scalars constructed from the vectors $A^i_M$, the {\it{free scalars}}, while the scalars coming from $\phi$, $g_{MN}$, and $B_{MN}$ are {\it{coupled scalars}}.  Since we will be seeking vacuum solutions in which $\sigma^\alpha=0$ and $A^i_M=0$, we can essentially drop those fields from the rest of the analysis.

The equations of motion for $\phi$, $g_{MN}$ and $B_{MN}$ are then (setting the free fields to zero)
\begin{align}
    \label{eq:PhiEOM}
    0=\ & \nabla^M\nabla_M\phi+\frac{1}{4}R-\frac{1}{48}\left|H_3\right|^2-\nabla^M\phi\nabla_M\phi,\\
    \label{eq:gEOM}
    0=\ & \hlf g_{MN}\left(R+4\nabla^P\nabla_P\phi-4\left(\partial\phi\right)^2-\frac{1}{12}\left|H_3\right|^2-e^{2\phi}\Lambda\right)-R_{MN}-2\nabla_M\nabla_N\phi+\frac{1}{4}H_M{}^{PR}H_{NPR},\\
    \label{eq:BEOM}
    0=\ & \nabla^P\left(e^{-2\phi}H_{MNP}\right).
\end{align}
Before we study the $AdS_{3}\times S^{3}$ solutions of these equations, let's show that for the 6D action in \eqref{6D string frame action}, the argument for the no-go theorem in \cite{Basile:2020mpt} can be repeated in the presence of an electrical flux to rule out deSitter. For this purpose, we have to write the action \eqref{6D string frame action} in the Einstein frame by using the transformation $g_{MN}\rr e^{\phi}g_{MN}$ and assuming a compactification into a 3D external spacetime and a 3D internal space with the following line element
$$
g_{MN}dx^{M}dx^{N}=g_{\m\n}dx^{\m}dx^{\n}+\tilde{g}_{ab}dy^{a}dy^{b}
$$
The dilaton and Einstein equations in the Einstein frame can be worked out to give the following
\begin{align}
\label{eq:dil Ein frame}
\tilde{\square}\phi=\ & \frac{3\Lambda}{2}e^{3\phi}-\frac{e^{-2\phi}}{12}|H_{3}|^{2}\\
\label{eq:Ein Ein frame}
R_{MN}=\ & \p_{M}\phi\p_{N}\phi+\frac{e^{-2\phi}}{4}\lp |H_{3}|^{2}_{MN}-\frac{1}{6}g_{MN}|H_{3}|^{2}\rp+\frac{\Lambda e^{3\phi}}{4}g_{MN}
\end{align}
where we have used the fact that the background dilaton field can only depend on the internal coordinates if we want the external space to have Lorentz invariance. Now, if we assume that the external spacetime is a maximally symmetric 3D spacetime, which obeys $R_{\m\n}=\Om g_{\m\n}$ where $\Om$ is the cosmological constant, and integrate \eqref{eq:dil Ein frame} and the trace of \eqref{eq:dil Ein frame} over the internal space, we arrive at the following conclusion
\be\label{no go result}
\int \Om=-\frac{1}{36}\int e^{-2\phi}|H_{3}|^{2}+\frac{1}{24}\int e^{-2\phi}|H_{3}|^{2}_{\m\n}g^{\m\n}
\ee
where the integrals are on the internal space and $|H_{3}|^{2}_{\m\n}g^{\m\n}=H_{\m MN}H^{\m MN}$. To maintain Lorentz invariance, $H_{\m MN}H^{\m MN}$ should be equal to $H_{\m \rho\s}H^{\m \rho\s}$ and thus, the second term on the right-hand side of \eqref{no go result} is zero if there is no electric flux. The absence of an electric flux implies that $\Om$ is negative, which is in accord with the no-go theorem in \cite{Basile:2020mpt}. The second term can be shown to be non-positive because for a 3-form like $H_{3}$ over the external spacetime has to be of the form $H_{\m\n\s}=k \e_{\m\n\s}$ where $k$ is a constant and therefore, the second term on the right hand side of \eqref{no go result} is
$$
-\frac{k^{2}}{4}\int e^{-2\phi}
$$
making it non-positive (and vanishing only when $k=0$, which corresponds to zero electric flux). Hence, \eqref{no go result} implies that $\Om$ should be negative even when there is an electric flux, which rules out deSitter.

Now, we look for vacuum solutions where $\phi=0$, the metric is that of $AdS_3\times S^3$ with scales $L_{AdS,\circ}$ and $L_\circ$ for $AdS_3$ and $S^3$ respectively, and with a background $H$-flux
\begin{equation}
    H_3=\frac{2}{\mathcal{L}_{AdS,\circ}}\epsilon_3+\frac{2}{\mathcal{L}_\circ}\epsilon_{S^3},
\end{equation}
where $\epsilon_3$ and $\epsilon_{S^3}$ are the volume forms for $AdS_3$ of radius $L_{AdS,\circ}$ and $S^3$ of radius $L_\circ$.  In particular this implies that
\begin{equation}
    \left|H_3\right|^2=H^{MNP}H_{MNP}=-\frac{24}{\mathcal{L}_{AdS,\circ}^2}+\frac{24}{\mathcal{L}_\circ^{2}},
\end{equation}
\begin{align}
    R_{\mu\nu}=\ & -\frac{2}{L_{AdS,\circ}^2}g_{\mu\nu},\\
    R_{\mu a}=\ & 0,\\
    R_{ab}=\ & \frac{2}{L_{\circ}^2}g_{ab},
\end{align}
and
\begin{equation}
    R=-\frac{6}{L_{AdS,\circ}^2}+\frac{6}{L_\circ^2},
\end{equation}
where we now use Greek indices for the $AdS_3$ directions and latin indices for the $S^3$ directions.  With this ansatz the $B$-field equations and block off-diagonal metric equations are satisfied.  The dilaton equation simply implies that $R=\frac{1}{12}|H_3|^2$, and the metric equations within the $AdS_3$ and $S^3$ blocks are satisfied provided that
\begin{equation}
    0=-\hlf\Lambda+\frac{2}{L_{AdS,\circ}^2}-\frac{2}{\mathcal{L}_{AdS,\circ}^2},\qquad 0=-\hlf\Lambda-\frac{2}{L_\circ^2}+\frac{2}{\mathcal{L}_\circ^2}.
\end{equation}
These equations let us rewrite all the length scales in terms of $L_\circ$,
\begin{equation}
    L_{AdS,\circ}^{-2}=L_\circ^{-2}-\frac{1}{4}\Lambda,\qquad\mathcal{L}_\circ^{-2}=L_\circ^{-2}+\frac{1}{4}\Lambda,\qquad\mathcal{L}_{AdS,\circ}^{-2}=L_\circ^{-2}-\hlf\Lambda.
\end{equation}
This also implies that $R=\frac{3}{2}\Lambda$, $|H_3|^2=18\Lambda$ which is consistent with what we found earlier.

Of course as we saw in section~\ref{sec:TreeLevel}, quantization fixes $\mathcal{L}_{AdS,\circ}$ and $\mathcal{L}_\circ$ in terms of the flux numbers $n_1$ and $n_5$ respectively, as well as $\alpha'$ and $g_\circ$ (and the volume $v$ of the four-torus). The relevant expressions are as follows
\be
\mathcal{L}_{AdS,\circ}=\frac{L^{3}_{\circ}v}{16\pi^{4}\al'g^{2}_{\circ}n_{1}}
\ee
\be
\mathcal{L}_{\circ}=\frac{L^{3}_{\circ}}{\al'n_{5}}
\ee
Therefore, the above relations allow us to determine $L_\circ$ and $g_\circ$ in terms of $n_1$ and $n_5$.  However, for the analysis in this section it is more convenient to keep everything in terms of $L_\circ$ and\footnote{This six-dimensional cosmological constant should not be confused with the three-dimensional cosmological constant given by $\La_{3,o}=-L_{AdS,o}^{-2}=-(L_o^{-2}-\frac{1}{4}\La)$.} $\Lambda=2\la g_\circ^2/\alpha'$.

\subsection{Torus moduli}

The one-loop contribution to the scalar potential is~\eqref{one loop potential}, which includes a factor of the integrated genus-one world-sheet partition function,
\be
V_{\text{1-loop}}=-\frac{\pi g_s^2}{\al'v}\int_{\mathcal{F}}\frac{d^2\tau}{\tau_2^2}Z(\tau;\s).
\ee
Here we have indicated explicitly that the partition function depends on some moduli $\s^\al$.  For a $T^4$ compactification, these are the eighty moduli of the $(20,4)$ signature lattice of the Narain compactification, consisting of ten metric moduli, six $B$-field periods, and $4\times 16=64$ Wilson line moduli.  Note that although the dimensionless torus volume $v$ also depends on the moduli $\s^\al$, the combination $g_o^2/v$ does not (that combination is fixed purely in terms of the fluxes), so the true $\s^\al$ dependence is entirely through the partition function.

In order to have a vacuum solution, we must sit at a critical point of this potential in the Narain moduli space.  In appendix~\ref{Appendix D} we show that for the square torus at the self-dual radius, all the first derivatives of the potential with respect to $\s^\al$ vanish (in fact we show the stronger result that the first derivatives of the partition function $Z(\tau;\s^\al)$ vanish even before integrating over the fundamental domain).  In a forthcoming article we will argue for the existence of several other critical points and compute both the value of the potential (i.e.~$\lambda$) as well as the Hessian matrix which determined the masses of the fields $\s^\al$ for each of them, as well as checking for so-called {\it{knife edge}} solutions~\cite{Ginsparg:1986wr,Fraiman:2023cpa}.  In the present work we will perform a less quantitative analysis.

The masses of the torus moduli will then depend on finding the eigenvalues of the Hessian at the critical point, but will certainly be of the form
\be
m_\al^2=\frac{g_o^2}{\al'v}\m_\al,
\ee
where $\m_\al$ is a dimensionless number, independent of the fluxes $n_1$ and $n_5$.  Although we do not expect that our critical points will be true minima of the scalar potential, meaning that at least some of the $\m_\al$ numbers will be negative numbers, the important question is whether the squared masses land above or below the BF bound, i.e.~whether
\be
L_{AdS,o}^2m_\al^2=\frac{L_{AdS,o}^2g_o^2}{\al'v}\m_\al\ge -1.
\ee
Near tree-level, this condition becomes
\be
s\frac{\m_\al}{\la}\ge -(2\pi)^4,
\ee
so for sufficiently small values $s$, all the torus moduli scalars will be above the BF bound (we're assuming here that our critical point is not a knife edge for which some directions are automatically unstable, regardless of the Hessian eigenvalues).  For the large $s$ limit, the condition becomes
\be
-1\le\lim_{s\rr\infty}L_{AdS,o}^2m_\al^2=2\sqrt{\frac{2}{3}}|n_5|\al'\sqrt{\frac{3}{2}}\frac{\m_\al}{\al'v\la|n_5|}\quad\Rightarrow\quad 2\frac{\m_\al}{\la}\ge -v.
\ee
So in this case, stability depends on the precise value of the most negative eigenvalue $\m_\al$.  Preliminary investigation seems to indicate that, at least for the square torus and some other critical points, that this inequality is violated for some eigenvalue $\m_\al$, and hence the intrinsically quantum solution is not perturbatively stable; rather there is some maximum value of $s$, or equivalently, minimum value of $n_1$, beyond which an instability develops.  In subsequent work we will explore this behavior in more detail.

\subsection{Perturbations}

Now we consider fluctuations around this vacuum solution.  We need to expand the fluctuations of the six-dimensional gravitational fields.  Besides the dilaton $\phi$, we have the metric and the $B$ field.  For the metric we write
\begin{equation}
    \d g_{\mu\nu}=H_{\mu\nu}+Mg_{\mu\nu},\qquad\d g_{\mu a}=S_{\mu a},\qquad\d g_{ab}=K_{ab}+Ng_{ab},
\end{equation}
where $H_{\mu\nu}$ and $K_{ab}$ are traceless, $g^{\mu\nu}H_{\mu\nu}=g^{ab}K_{ab}=0$.  For the $B$ field we take
\begin{equation}
    \d B_{\mu\nu}=\epsilon_{\mu\nu\rho}U^\rho,\qquad\d B_{\mu a}=C_{\mu a},\qquad\d B_{ab}=\epsilon_{abc}V^c,
\end{equation}
where $\epsilon_{\mu\nu\rho}$ and $\epsilon_{abc}$ are the antisymmetric tensors (i.e.~volume forms) on $AdS_3$ and $S^3$ respectively.

We will impose a Donder-Lorentz-like gauge on the diffeomorphisms and $B$-field gauge transformations, demanding that
\begin{equation}
    \label{eq:DonderLorentz}
    \nabla^aS_{\mu a}=0,\qquad\nabla^bK_{ab}=0,\qquad\nabla^aC_{\mu a}=0,\qquad\nabla_{[a}V_{b]}=0.
\end{equation}

Next we expand each of our equations of motion, (\ref{eq:PhiEOM}), (\ref{eq:gEOM}), and (\ref{eq:BEOM}), to linear order in the fluctuating fields, using the gauge condition to simplify slightly.  Collecting the results together and using notation $\Box_0=\nabla^\m\nabla_\m$, $\Box_x=\nabla^a\nabla_a$ for the Laplacians on AdS$_3$ and $S^3$ respectively, we find the following.

\textbf{Dilaton}:
\begin{equation}
\label{dilaton}
    0=\frac{1}{4}\Box_{0}\lp 2M+3N-4\phi\rp+\frac{1}{4}\Box_{x}\lp 3M+2N-4\phi\rp-\frac{3\Lambda}{8}(M+N)-\frac{1}{4}\nabla_{\m}\nabla_{\n}H^{\m\n}+\frac{1}{2\mathcal{L}_{o}}\nabla_{a}V^{a}-\frac{1}{2\mathcal{L}_{AdS,o}}\nabla_{\m}U^{\m},
\end{equation}
\textbf{$g_{\mu\nu}$ Trace component}:
\begin{align}
    \label{g mu nu trace}
    0=\ & \frac{12M}{\mathcal{L}^{2}_{AdS,o}}+\frac{3\Lambda}{4}\lp M+3N-4\phi\rp-\frac{3}{\mathcal{L}_{AdS,o}}\nabla_{\m}U^{\m}-\Box_{0}(M+3N-4\phi)\non\\
    & \quad +\hlf \nabla_{\m}\nabla_{\n}H^{\m\n}-\frac{3}{\mathcal{L}_{o}}\nabla_{a}V^{a}-3\Box_{x}\lp M+N-2\phi\rp,
\end{align}
\textbf{$g_{\mu\nu}$ Traceless component}:
\begin{align}
    \label{g mu nu traceless}
    0=\ & \frac{4}{L^{2}_{AdS,o}} H_{\m\n}-\nabla_{\m}\nabla_{\n}\lp M+3N-4\phi\rp+2\nabla_{\rho}\nabla_{(\m}H_{\n)}^{\;\;\rho}-(\Box_{0}+\Box_{x})H_{\m\n}\non\\
    & \quad +\frac{1}{3}g_{\m\n}\Box_{0}\lp M+3N-4\phi\rp-\frac{2}{3}g_{\m\n}\nabla_{\s}\nabla_{\rho}H^{\s\rho},
\end{align}
\textbf{$g_{ab}$ Trace component}:
\begin{align}
    \label{g ab trace}
    0=\ & -4L^{-2}_{o}N+\frac{\Lambda}{4}\lp 3M-3N-4\phi\rp+\mathcal{L}^{-1}_{AdS,o}\nabla_{\m}U^{\m}-\Box_{0}\lp M+N-2\phi\rp+\hlf\nabla_{\m}\nabla_{\n}H^{\m\n}\non\\
    & \quad +\mathcal{L}^{-1}_{o}\nabla_{a}V^{a}-\frac{1}{3}\Box_{x}\lp 3M+N-4\phi\rp,
\end{align}
\textbf{$g_{ab}$ Traceless component}:
\begin{equation}
    \label{g ab traceless}
    0=2\nabla_{\m}\nabla_{(a}S^{\m}{}_{b)}-\lp\Box_{0}+\Box_{x}-2L^{-2}_{o}\rp K_{ab}-\nabla_{\{a}\nabla_{b\}}\lp 3M+N-4\phi\rp,
\end{equation}
\textbf{$g_{\m a}$ component}:
\begin{align}
    \label{g mu a}
    0=\ & \frac{3\Lambda}{2} S_{\m a}-\lp \Box_{0}+\Box_{x}\rp S_{\m a}+\nabla_{\n}\nabla_{a}H_{\m}^{\;\;\n}-2\mathcal{L}^{-1}_{o}\;\nabla_{\m}V_{a}+\nabla_{\m}\nabla_{\n}S^{\n}{}_{a}-2\nabla_{\m}\nabla_{a}\lp M+N-2\phi\rp\non\\
    & \quad +2\mathcal{L}^{-1}_{AdS,o}\lp \e_{\m\n\rho}\nabla^{\rho}C^{\n}_{\;\;a}+\nabla_{a}U_{\m}\rp-2\mathcal{L}^{-1}_{o}\;\e_{abc}\nabla^{c}C_{\m}^{\;\;b},
\end{align}
\textbf{$B_{\m \n}$ component}:
\begin{equation}
    \label{B mu nu}
    0=\mathcal{L}^{-1}_{AdS,o}\nabla^{\m}\lp 3M-3N+4\phi\rp-\nabla^{\m}\nabla_{\n}U^{\n}-\Box_{x}U^{\m},
\end{equation}
\textbf{$B_{ab}$ component}:
\begin{equation}
    \label{B ab}
    0=2\mathcal{L}^{-1}_{o}\nabla_{\m}S^{\m c}-\Box_{0}V^{c}-\e_{\;\;ab}^{c}\nabla_{\m}\nabla^{b}C^{\m a}-\mathcal{L}^{-1}_{o}\nabla^{c}\lp 3M-3N-4\phi\rp-\nabla^{c}\nabla_{a}V^{a},
\end{equation}
\textbf{$B_{\m a}$ component}:
\begin{equation}
    \label{B mu a}
    0=\lp \frac{\Lambda}{2}-\Box_{0}-\Box_{x}\rp C_{\m a}+\nabla_{\m}\nabla_{\n}C^{\n}_{\;\;a}+2\mathcal{L}^{-1}_{AdS,o}\e_{\m\lambda\n}\nabla^{\n}S^{\lambda}{}_{a}-\e_{\m\la\n}\nabla^{\n}\nabla_{a}U^{\la}-2\mathcal{L}^{-1}_{o}\e_{abc}\nabla^{c}S_{\m}{}^{b}.
\end{equation}

\subsection{Spherical harmonics}

Next we will expand each of these fields in spherical harmonics on $S^3$.  We recall that a scalar field on $S^3$ can be expanded in harmonics $Y^{(\ell,0)}$, $\ell\ge 0$, vector fields can be expanded in terms of $Y^{(\ell,\pm 1)}_a$ and $\nabla_aY^{(\ell,0)}$, $\ell\ge 1$, and symmetric traceless two-tensors can be expanded in terms of $Y^{(\ell,\pm 2)}_{ab}$, $\nabla_{\{a}Y^{(\ell,\pm 1)}_{b\}}$, and $\nabla_{\{a}\nabla_{b\}}Y^{(\ell,0)}$, $\ell\ge 2$.  We have also introduced notation for traceless symmetrization,
\begin{equation}
    T_{\{ab\}}:=\hlf T_{ab}+\hlf T_{ba}-\frac{1}{3}g_{ab}T^c{}_c.
\end{equation}
The spherical harmonics satisfy some identities,
\begin{align}
    \square_xY^{(\ell,0)}=\ & -L_\circ^{-2}\ell\lp\ell+2\rp Y^{(\ell,0)},\\
    \nabla^aY^{(\ell,\pm 1)}_a=\ & 0,\\
    \square_xY^{(\ell,\pm 1)}_a=\ & -L_\circ^{-2}\lp\ell^2+2\ell-1\rp Y^{(\ell,\pm 1)}_a,\\
    \epsilon_a{}^{bc}\nabla_bY^{(\ell,\pm 1)}_c=\ & \pm L_\circ^{-1}\lp\ell+1\rp Y^{(\ell,\pm 1)}_a,\\
    g^{ab}Y^{(\ell,\pm 2)}_{ab}=\ & 0,\\
    \nabla^bY^{(\ell,\pm 2)}_{ab}=\ & 0,\\
    \square_xY^{(\ell,\pm 2)}_{ab}=\ & -L_\circ^{-2}\lp\ell^2+2\ell-2\rp Y^{(\ell,\pm 2)}_{ab}.
\end{align}
We use $\square_x=\nabla^a\nabla_a$ to denote the Laplacian on $S^3$ and $\square_0=\nabla^\mu\nabla_\mu$ to denote the Laplacian on $AdS_3$.  We also have low $\ell$ properties
\begin{equation}
    \nabla_aY^{(0,0)}=0,\qquad\nabla_{\{a}\nabla_{b\}}Y^{(1,0)}=0,\qquad\nabla_{(a}Y^{(1,\pm 1)}_{b)}=0.
\end{equation}
We will expand each of our fields in this basis, e.g.~
\begin{align}
    \phi=\ & \sum_{\ell=0}^\infty\phi^{(\ell,0)}Y^{(\ell,0)},\\
    S_{\mu a}=\ & \sum_{\ell=1}^\infty\lp S_\mu^{(\ell,1)}Y^{(\ell,1)}_a+S_\mu^{(\ell,-1)}Y^{(\ell,-1)}_a+S_\mu^{(\ell,0)}\nabla_aY^{(\ell,0)}\rp,\\
    K_{ab}=\ & \sum_{\ell=2}^\infty\lp K^{(\ell,2)}Y^{(\ell,2)}_{ab}+K^{(\ell,-2)}Y^{(\ell,-2)}_{ab}+K^{(\ell,1)}\nabla_{(a}Y^{(\ell,1)}_{b)}+K^{(\ell,-1)}\nabla_{(a}Y^{(\ell,-1)}_{b)}+K^{(\ell,0)}\nabla_{\{a}\nabla_{b\}}Y^{(\ell,0)}\rp.
\end{align}
Note, however, that once we impose our gauge conditions (\ref{eq:DonderLorentz}), some of the terms in this expansion will go away, specifically we will have $S_\m^{(\ell,0)}=0$ and $K^{(\ell,\pm 1)}=K^{(\ell,0)}=0$, $C_\m^{(\ell,0)}=0$, and $V^{(\ell,\pm 1)}=0$.

\subsection{Residual gauge transformations}

We can also consider six-dimensional diffeomorphisms with parameter $\xi^M$ and six-dimensional $B$-field gauge transformations with parameters $\lambda_M$.  We will work only to first order in fluctuations or gauge parameters.  Expanding in spherical harmonics the diffeomorphism parameters have components
\begin{align}
    \xi_\mu=\ & \sum_{\ell=0}^\infty\xi_\mu^{(\ell,0)}Y^{(\ell,0)},\\
    \xi_a=\ & \sum_{\ell=1}^\infty\left(\xi^{(\ell,1)}Y^{(\ell,1)}_a+\xi^{(\ell,-1)}Y^{(\ell,-1)}_a+\xi^{(\ell,0)}\nabla_aY^{(\ell,0)}\right).
\end{align}

Under these gauge transformations we have in particular
\begin{equation}
    \d S_{\mu a}=\nabla_\mu\xi_a+\nabla_a\xi_\mu=\sum_{\ell=1}^\infty\left(\nabla_\mu\xi^{(\ell,1)}Y^{(\ell,1)}_a+\nabla_\mu\xi^{(\ell,-1)}Y^{(\ell,-1)}_a+\left(\nabla_\mu\xi^{(\ell,0)}+\xi_\mu^{(\ell,0)}\right)\nabla_aY^{(\ell,0)}\right),
\end{equation}
so to preserve the gauge $S_\mu^{(\ell,0)}=0$, we must have $\xi_\mu^{(\ell,0)}=-\nabla_\mu\xi^{(\ell,0)}$, $\ell\ge 1$.  Similarly,
\begin{equation}
    \d K_{ab}=2\nabla_{\{a}\xi_{b\}}=2\sum_{\ell=2}^\infty\left(\xi^{(\ell,1)}\nabla_{\{a}Y^{(\ell,1)}_{b\}}+\xi^{(\ell,-1)}\nabla_{\{a}Y^{\ell,-1)}_{b\}}+\xi^{(\ell,0)}\nabla_{\{a}\nabla_{b\}}Y^{(\ell,0)}\right),
\end{equation}
which preserves our chosen gauge only if $\xi^{(\ell,1)}=\xi^{(\ell,-1)}=\xi^{(\ell,0)}=0$ for $\ell\ge 2$.  The only surviving diffeomorphisms are
\begin{equation}
    \xi_\mu=\xi_\mu^{(0,0)}Y^{(0,0)}-\nabla_\mu\xi^{(1,0)}Y^{(1,0)},\qquad\xi_a=\xi^{(1,1)}Y^{(1,1)}_a+\xi^{(1,-1)}Y^{(1,-1)}_a+\xi^{(1,0)}\nabla_aY^{(1,0)}.
\end{equation}

Turning to the $B$-fields, we'll focus first on the $B$-field gauge transformations with parameters
\begin{align}
    \lambda_\mu=\ & \sum_{\ell=0}^\infty\lambda_\mu^{(\ell,0)}Y^{(\ell,0)},\\
    \lambda_a=\ & \sum_{\ell=1}^\infty\left(\lambda^{(\ell,1)}Y^{(\ell,1)}_a+\lambda^{(\ell,-1)}Y^{(\ell,-1)}_a+\lambda^{(\ell,0)}\nabla_aY^{\ell,0)}\right).
\end{align}
Under these transformations we have
\begin{equation}
    \d C_{\mu a}=\nabla_\mu\lambda_a-\nabla_a\lambda_\mu=\sum_{\ell=1}^\infty\left(\nabla_\mu\lambda^{(\ell,1)}Y^{(\ell,1)}_a+\nabla_\mu\lambda^{(\ell,-1)}Y^{(\ell,-1)}_a+\left(\nabla_\mu\lambda^{(\ell,0)}-\lambda^{(\ell,0)}_\mu\right)\nabla_aY^{(\ell,0)}\right).
\end{equation}
Demanding that $\d C^{(\ell,0)}_\mu=0$ then requires that $\lambda_\mu^{(\ell,0)}=\nabla_\mu\lambda^{(\ell,0)}$ for $\ell\ge 1$, but this just means that the $(\ell,0)$ parts of the gauge transformation comprise an exact one-form and so drop out of all gauge transformations.  We will set $\lambda_\mu^{(\ell,0)}$ and $\lambda^{(\ell,0)}$ to zero for $\ell\ge 1$ henceforth.

Similarly,
\begin{equation}
    \d V^a=\epsilon^{abc}\nabla_b\lambda_c=L_\circ^{-1}\sum_{\ell=1}\left(\ell+1\right)\left(\lambda^{(\ell,1)}Y^{(\ell,-1)}_a-\lambda^{(\ell,-1)}Y^{(\ell,-1)}_a\right),
\end{equation}
so preserving the gauge condition that $V^{(\ell,\pm 1)}=0$ requires $\lambda^{(\ell,\pm 1)}=0$ as well, for $\ell\ge 1$.  The only residual $B$-field gauge transformations are the $(0,0)$ ones, $\lambda_\mu=\lambda_\mu^{(0,0)}Y^{(0,0)}$.

The $B$-fields also have vacuum expectation values (corresponding to the $H$-fluxes that wrap AdS$_3$ and $S^3$), so the fluctuations will also transform under the residual diffeomorphisms found above.  In order to preserve the gauge conditions on these fields, the diffeomorphisms must be paired with compensating $B$-field gauge transformations.  The final form of the $B$-field fluctuation transformations is most easily determined by demanding that the equations of motion remain invariant.

Putting everything together, these residual transformations have non-trivial actions on the following fields,
\begin{align}
    \d S_\mu^{(1,\pm 1)}=\ & \nabla_\mu\xi^{(1,\pm 1)},\\
    \d C_\mu^{(1,\pm 1)}=\ & \mp L_\circ\mathcal{L}_\circ^{-1}\nabla_\mu\xi^{(1,\pm 1)},
\end{align}
\begin{align}
    \d H_{\mu\nu}^{(1,0)}=\ & -2\nabla_\mu\nabla_\nu\xi^{(1,0)}+\frac{2}{3}g_{\mu\nu}\square_0\xi^{(1,0)},\\
    \d M^{(1,0)}=\ & -\frac{2}{3}\square_0\xi^{(1,0)},\\
    \d N^{(1,0)}=\ & -2L_\circ^{-2}\xi^{(1,0)},\\
    \d V^{(1,0)}=\ & 2\mathcal{L}_\circ^{-1}\xi^{(1,0)},\\
    \d U_\mu^{(1,0)}=\ & -2\mathcal{L}_{AdS,\circ}^{-1}\nabla_\mu\xi^{(1,0)},
\end{align}
and
\begin{align}
    \d H_{\mu\nu}^{(0,0)}=\ & \nabla_\mu\xi_\nu^{(0,0)}+\nabla_\nu\xi_\mu^{(0,0)}-\frac{2}{3}g_{\mu\nu}\nabla^\rho\xi_\rho^{(0,0)},\\
    \d M^{(0,0)}=\ & \frac{2}{3}\nabla^\mu\xi_\mu^{(0,0)},\\
    \label{eq:U00Transformation}
    \d U_\mu^{(0,0)}=\ & 2\mathcal{L}_{AdS,\circ}^{-1}\xi_\mu^{(0,0)}+\epsilon_\mu{}^{\nu\rho}\nabla_\nu\lambda^{(0,0)}_\rho.
\end{align}

\subsection{Spectrum}

Plugging these expansions into the equations of motion, we can extract the coefficients of each spherical harmonic structure.

\begin{itemize}
\item $(\ell,\pm 2)$:
This structure only appears in the traceless $g_{ab}$ equation.  Reading off the coefficient of $Y^{(\ell,\pm 2)}_{ab}$, $\ell\ge 2$ gives the equation
\begin{equation}
    0=-\lp\square_0-L_\circ^{-2}\ell\lp\ell+2\rp\rp K^{(\ell,\pm 2)}.
\end{equation}
In other words, the modes of $K_{ab}$ give rise to scalar fields in three dimensions with masses $m^2=L_\circ^{-2}\ell\lp\ell+2\rp$, which are all positive.
\item $(\ell,\pm 1)$:
For $\ell\ge 2$, we get one equation from the coefficient of $\nabla_{(a}Y^{(\ell,\pm 1)}_{b)}$ in the traceless $g_{ab}$ equation,
\begin{equation}
    0=2\nabla^\mu S^{(\ell,\pm 1)}_\mu,
\end{equation}
while the coefficients of $Y^{(\ell,\pm 1)}_a$, $\ell\ge 1$, in the $g_{\mu a}$, $B_{ab}$, and $B_{\mu a}$ equations give,
\begin{align}
    0=\ & -\square_0S^{(\ell,\pm 1)}_\mu+\nabla_\mu\nabla^\nu S^{(\ell,\pm 1)}_\nu-2\mathcal{L}_{AdS,\circ}^{-1}\epsilon_\mu{}^{\nu\rho}\nabla_\nu C^{(\ell,\pm 1)}_\rho\non\\
    & \quad +\lp L_\circ^{-2}\lp\ell^2+2\ell-1\rp+\frac{3}{2}\Lambda\rp S^{(\ell,\pm 1)}_\mu\pm 2L_\circ^{-1}\mathcal{L}_\circ^{-1}\lp\ell+1\rp C^{(\ell,\pm 1)}_\mu,\\
    0=\ & 2\mathcal{L}_\circ^{-1}\nabla^\mu S^{(\ell,\pm 1)}_\mu\pm L_\circ^{-1}\lp\ell+1\rp\nabla^\mu C^{(\ell,\pm 1)}_\mu,\\
    0=\ & -\square_0C^{(\ell,\pm 1)}_\mu+\nabla_\mu\nabla^\nu C^{(\ell,\pm 1)}_\nu-2\mathcal{L}_{AdS,\circ}^{-1}\epsilon_\mu{}^{\nu\rho}\nabla_\nu S^{(\ell,\pm 1)}_\rho\non\\
    & \quad +\lp L_\circ^{-2}\lp\ell^2+2\ell-1\rp+\frac{\Lambda}{2}\rp C^{(\ell,\pm 1)}_\mu\pm 2L_\circ^{-1}\mathcal{L}_\circ^{-1}\lp\ell+1\rp S^{(\ell,\pm 1)}_\mu.
\end{align}

So for $\ell\ge 2$ we have $\nabla^\mu S^{(\ell,\pm 1)}_\mu=\nabla^\mu C^{(\ell,\pm 1)}_\mu=0$, and remaining equations
\begin{align}
\square_0S_\mu^{(\ell,\pm 1)}=\ & -2\mathcal{L}_{AdS,\circ}^{-1}\epsilon_\mu{}^{\nu\rho}\nabla_\nu C_\rho^{(\ell,\pm 1)}+\lp L_\circ^{-2}\lp\ell^2+2\ell-1\rp+\frac{3}{2}\Lambda\rp S_\mu^{(\ell,\pm 1)}\pm 2L_\circ^{-1}\mathcal{L}_\circ^{-1}\lp\ell+1\rp C_\mu^{(\ell,\pm 1)},\\
\square_0C_\mu^{(\ell,\pm 1)}=\ & -2\mathcal{L}_{AdS,\circ}^{-1}\epsilon_\mu{}^{\nu\rho}\nabla_\nu S_\rho^{(\ell,\pm 1)}+\lp L_\circ^{-2}\lp\ell^2+2\ell-1\rp+\hlf\Lambda\rp C_\mu^{(\ell,\pm 1)}\pm 2L_\circ^{-1}\mathcal{L}_\circ^{-1}\lp\ell+1\rp S_\mu^{(\ell,\pm 1)}.
\end{align}
For $\ell=1$ we don't have the equation that imposes $\nabla^\mu S_\mu^{(1,\pm 1)}$, but we do have a residual gauge transformation with parameter $\xi^{(1,\pm 1)}$ that can be used to impose this as a gauge condition, resulting in the same two equations as above but setting $\ell=1$.  It would be interesting to do a more detailed analysis of the stability implications of these vectors.

\item $(\ell,0)$ scalars: Let's start with the scalar fields in three dimensions.  We'll drop the $(\ell,0)$ superscripts on all fields, but it should be understood that we are only working with these modes.  Additionally, we will rescale the field $V$, defining\footnote{Not to be confused with the $\mathcal{V}$ appearing in~\eqref{Einstein frame def}.}
\be
\mathcal{V}=\mathcal{L}_\circ^{-1}V.
\ee

For $\ell\ge 2$ the $(\ell,0)$ piece of the traceless $g_{ab}$ equation tells us that
\be
3M+N-4\phi=0,
\ee
which can be used to eliminate $\phi$.  The $(\ell,0)$ part of the $B_{\mu a}$ equation tells us that the one-form $U_\mu$ is closed in AdS$_3$, so we can locally write it as a gradient of a scalar field $u$, $U_\mu=\mathcal{L}_{AdS,\circ}\nabla_\mu u$.  Of course, this $u$ is only defined up to constant shifts.  Finally, by taking the divergence of the $(\ell,0)$ part of the $g_{\m a}$ equation we get
\be
\nabla^\m\nabla^\n H_{\m\n}=\square_0\lp -M+N+2\mathcal{V}-2u\rp.
\ee
Substituting these into the equations for the dilaton, the $g_{\m\n}$ trace, and the $g_{ab}$ trace respectively, we get the following equations
\begin{align}
0=\ & \square_0\lp\frac{1}{4}N-\hlf\mathcal{V}\rp+L_\circ^{-2}\ell\lp\ell+2\rp\lp -\frac{1}{4}N-\hlf\mathcal{V}\rp+\Lambda\lp -\frac{3}{8}M-\frac{3}{8}N\rp,\\
0=\ & \square_0\lp\frac{3}{2}M-\frac{3}{2}N+\mathcal{V}-4u\rp+12L_\circ^{-2}M-L_\circ^{-2}\ell\lp\ell+2\rp\lp\frac{3}{2}M-\frac{3}{2}N-3\mathcal{V}\rp+\Lambda\lp -\frac{15}{2}M+\frac{3}{2}N\rp,\\
0=\ & \square_0\mathcal{V}-4L_\circ^{-2}N-L_\circ^{-2}\ell\lp\ell+2\rp\mathcal{V}-\Lambda N,
\end{align}
The $(\ell,0)$ part of the $B_{ab}$ equation is redundant, being equivalent to the $g_{ab}$ trace equation above.  The $B_{\m\n}$ equation becomes
\be
0=\nabla_\m\ls -\square_0u+L_\circ^{-2}\lp 6M-2N\rp+L_\circ^{-2}\ell\lp\ell+2\rp u+\Lambda\lp -3M+N\rp\rs.
\ee
The solution here is that the quantity in square brackets must be a constant, and we can use the constant shifts in $u$ to set that constant to zero.  Combining this with the three equations above, we can rearrange to get the mass matrix,
\be
\square_0\lp\begin{matrix} M \\ N \\ \mathcal{V} \\ u \end{matrix}\rp=\lp\begin{matrix} 8a+b-\frac{3}{2}\La & \frac{9}{2}\La & \frac{4}{3}b & \frac{8}{3}b \\ \frac{3}{2}\La & 8a+b+\frac{7}{2}\La & 4b & 0 \\ 0 & 4a+\La & b & 0 \\ 6a-3\La & -2a+\La & 0 & b \end{matrix}\rp\lp\begin{matrix} M \\ N \\ \mathcal{V} \\ u \end{matrix}\rp,
\ee
where we have defined,
\be
a=L_\circ^{-2},\qquad b=L_\circ^{-2}\ell\lp\ell+2\rp.
\ee

If we work to tree level, setting $\La=0$ and $L_\circ=L$, then the eigenvalues of this mass matrix are
\be
m^2_1=m^2_2=L^{-2}\ell\lp\ell-2\rp,\qquad m^2_3=m^2_4=L^{-2}\lp\ell+2\rp\lp\ell+4\rp,
\ee
which for $\ell\ge 2$ are all non-negative and hence well above the Breitenlohner-Friedmann bound of $-L^{-2}$.  Once we incorporate one loop effects we can either solve for the eigenvalues numerically, or we can write the eigenvalues as expansions,
\begin{align}
\label{eq:M12Eigenvalues}
L_{AdS,\circ}^2m^2_{1,2}=\ & \ell\lp\ell-2\rp+\frac{\la g_s^2L^2\ell}{2\al'\lp\ell+1\rp}\lp\ell^2+2\pm\sqrt{9\ell^2+16}\rp+\mathcal{O}(\la^2),\\
L_{AdS,\circ}^2m^2_{3,4}=\ & \lp\ell+2\rp\lp\ell+4\rp+\frac{\la g_s^2L^2\lp\ell+2\rp}{2\al'\lp\ell+1\rp}\lp\ell^2+4\ell+6\pm\sqrt{9\ell^2+36\ell+52}\rp+\mathcal{O}(\la^2).
\end{align}
Here the expansions have been written so as to easily compare with the Brightenlohner-Friedmann bound of $m_{BF}^2=-L_{AdS,\circ}^{-2}$.  The quantities on the right-hand side are the tree-level values ($L^2=\al'|n_5|$, $g_s^2=(v/(2\pi)^4)(n_5/n_1)$).  The full numerical results for a range of fluxes are plotted in Figure~\ref{fig:l=2 eigenvalues fixed n1} for $\ell=2$.  Note that the combinations $L_{AdS,\circ}^2m^2_i$ only depend on the fluxes $n_1$ and $n_5$ through the combination $n_5^2/n_1$, so it's enough to fix one of the fluxes (say $n_1$) and vary the other.  For $\ell=2$, one of the eigenvalues is negative, but remains above the BF bound, while for $\ell>2$ all four eigenvalues are positive.  We note that the numerical solutions for $L_{AdS,\circ}^2m^2$ asymptote to constant values both in the small $n_5$ limit, where we recover the tree level results $\ell(\ell-2)$ and $(\ell+2)(\ell+4)$, but also in the large $n_5$ limit.  This is of course equivalent to the $n_1\rightarrow 0$ limit discussed previously, and will be examined more closely in section \ref{subsec:Smalln1}.  

\begin{figure}
    \centering
    \includegraphics[width=0.5\linewidth]{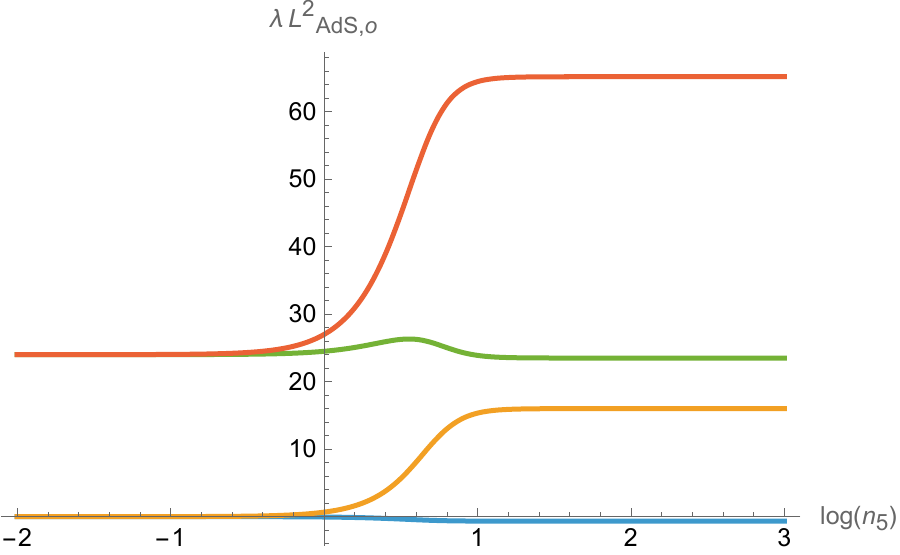}
    \caption{The four eigenvalues for the $\ell=2$ mass matrix plotted against $\log n_{5}$ for $n_{1}=10$.  On the left of the diagram (corresponding to small values of $n_5^2/n_1$) the eigenvalues approach their tree-level, or supersymmetric, values for which $L_{AdS,o}^2m^2$ are $\ell(\ell-2)=0$ and $(\ell+2)(\ell+4)=24$ in this case.  To the right (large values of $n_5^2/n_1$) the eigenvalues split and asymptote to another set of values (see section \ref{subsec:Smalln1} for further discussion).  Although one eigenvalue is negative, it remains above the BF bound $L_{AdS,o}^2m_{BF}^2=-1$ for all values of the fluxes.  For $\ell>2$, all four eigenvalues are positive for the whole range of fluxes.}
    \label{fig:l=2 eigenvalues fixed n1}
\end{figure}

For $\ell=1$ the analysis is almost the same.  We no longer have the equation $3M+N-4\phi=0$ coming from the traceless $g_{ab}$ equation, since that was only valid for $\ell\ge 2$.  However, we do have residual gauge transformations parameterized by $\xi^{(1,0)}$.  We can use this freedom to impose $4\phi=3M+N$ as a gauge condition.  There are still some residual gauge transformations left after this step, namely ones for which
\be
0=3\d M+\d N-4\d\phi=-2\lp\square_0+L_\circ^{-2}\rp\xi^{(1,0)}.
\ee
Once this gauge condition has been imposed, we can follow the same steps as above and arrive at the same mass matrix, except with the condition that $b=3a$, since $\ell=1$.  In this particular case, it turns out that the mass matrix has one particularly simple eigenvalue, $m_1^2=-a=-L_\circ^{-2}$, which only receives loop corrections implicitly through $L_\circ$ (this corresponds to the eigenvalue with $\ell=1$ and the lower sign chosen in (\ref{eq:M12Eigenvalues})).  This is precisely the correct value to be removed by our residual gauge transformations.  The remaining three eigenvalues correspond to physical scalar modes.  They have expansions
\begin{align}
L_{AdS,\circ}^2m_1^2=\ & -1+\frac{2\la g_s^2L^2}{\al'}+\mathcal{O}(\la^2),\\
L_{AdS,\circ}^2m_{2,3}^2=\ & 15+\frac{3\la g_s^2L^2}{4\al'}\lp 11\pm\sqrt{97}\rp+\mathcal{O}(\la^2).
\end{align}
In particular, at tree level one of the masses is right at the BF bound, but the first order correction to $m^2$ is positive, and so doesn't lead to a perturbative instability.  Looking at the numerical solution, plotted in Figure \ref{fig:l=1 eigenvalues fixed n1}, we can see that the eigenvalue indeed remains above the BF bound for all values of the fluxes.

\begin{figure}
    \centering
    \includegraphics[width=0.5\linewidth]{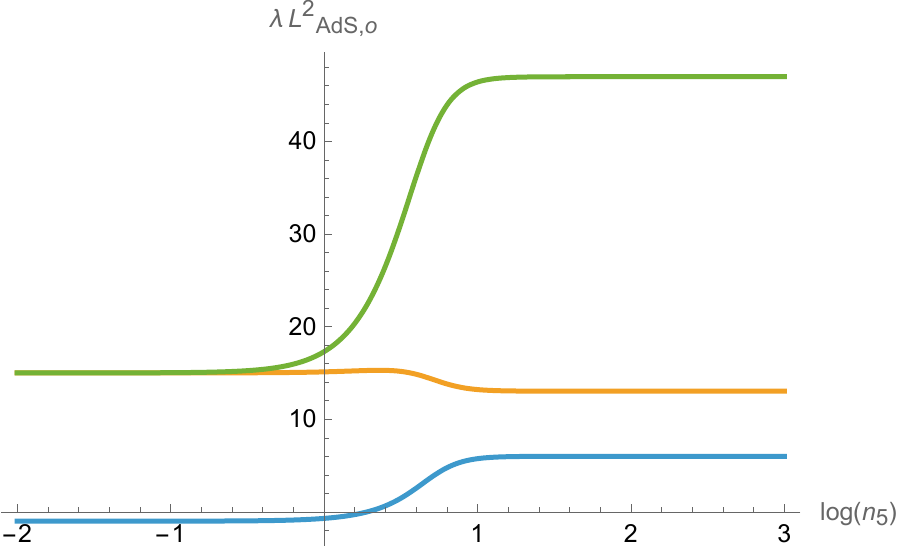}
    \caption{The three eigenvalues for the $\ell=1$ mass matrix plotted against $\log n_{5}$ for $n_{1}=10$.  At small $n_5$ we get the tree-level values, which are two eigenvalues at $L_{AdS,o}^2m^2=(\ell+2)(\ell+4)=15$, and one more at $\ell(\ell-2)=-1$, which is the BF bound.  For large $n_5$ we asymptote to different values.  All eigenvalues are above the BF bound for all values of the fluxes.}
    \label{fig:l=1 eigenvalues fixed n1}
\end{figure}

Finally, we turn to $\ell=0$.  In this case the field $\mathcal{V}$ is no longer present, and we also lose the traceless $g_{ab}$ equation, the $g_{\m a}$, the $B_{\m a}$ equation, and the $B_{ab}$ equation.  Although we have lost the equation which tells us that $U_\mu$ is a closed one-form, we can now impose that as a gauge condition using the residual $B$-field gauge transformations parameterized by $\la_\mu^{(0,0)}$.  This still leaves some residual gauge transformations, namely ones for which
\be
0=\d\lp\e_\m{}^{\n\rho}\nabla_\n U_\rho\rp=2\mathcal{L}_{AdS,\circ}^{-1}\e_\m{}^{\n\rho}\nabla_\n\xi_\rho^{(0,0)}+\lp\square_0+2L_{AdS,\circ}^{-2}\rp\la_\m^{(0,0)}.
\ee
In particular, any diffeomorphism generated by $\xi_\m^{(0,0)}=\nabla_\m\al$ will not disturb this condition.  Once we have imposed that $U_\m$ is closed, we can again write it locally as $U_\m=\mathcal{L}_{AdS,\circ}\nabla_\m u$, and we see from (\ref{eq:U00Transformation}) that under the gauge transformation parameterized by $\al$ above we have $\d u=2\mathcal{L}_{AdS,\circ}^{-2}\al$.  

Next we use the $\al$ gauge transformation to set
\be
M=N-\frac{4}{3}\phi.
\ee
We still have residual transformations from parameters $\al$ that satisfy $\square_0\al=0$.  Now substituting the above equation for $M$, the $B_{\m\n}$ equation becomes simply
\be
\square_0u=0,
\ee
but the residual $\al$ gauge transformation can then be precisely used to set $u=0$.  For the remaining dilaton, $g_{\m\n}$ trace, and $g_{ab}$ trace equations, we get
\begin{align}
0=\ & \square_0\lp\frac{5}{4}N-\frac{5}{3}\phi\rp+\La\lp -\frac{3}{4}N+\hlf\phi\rp-\frac{1}{4}\nabla^\m\nabla^\n H_{\m\n},\\
0=\ & \square_0\lp -4N+\frac{16}{3}\phi\rp+L_\circ^{-2}\lp 12N-16\phi\rp+\La\lp -3N+4\phi\rp+\hlf\nabla^\m\nabla^\n H_{\m\n},\\
0=\ & \square_0\lp -2N+\frac{10}{3}\phi\rp+L_\circ^{-2}\lp -4N\rp+\La\lp -2\phi\rp+\hlf\nabla^\m\nabla^\n H_{\m\n}.
\end{align}
Eliminating $\nabla^\m\nabla^\n H_{\m\n}$ and rearranging, we get the mass matrix
\begin{align}
\square_0N=\ & \lp 8L_\circ^{-2}+3\La\rp N+2\La\phi,\\
\square_0\phi=\ & \frac{9}{2}\La N+\lp 8L_\circ^{-2}-\La\rp\phi,
\end{align}
corresponding to eigenvalues
\be
m^2_{1,2}=8L_\circ^{-2}+\La\pm\sqrt{13}\La,
\ee
or
\be
L_{AdS,\circ}^2m^2_{1,2}=8+\frac{2\la g_s^2L^2}{\al'}\lp 3\pm\sqrt{13}\rp+\mathcal{O}(\la^2).
\ee
The exact values are plotted in Figure \ref{fig:l=0 eigenvalues fixed n1}.

\begin{figure}
    \centering
    \includegraphics[width=0.5\linewidth]{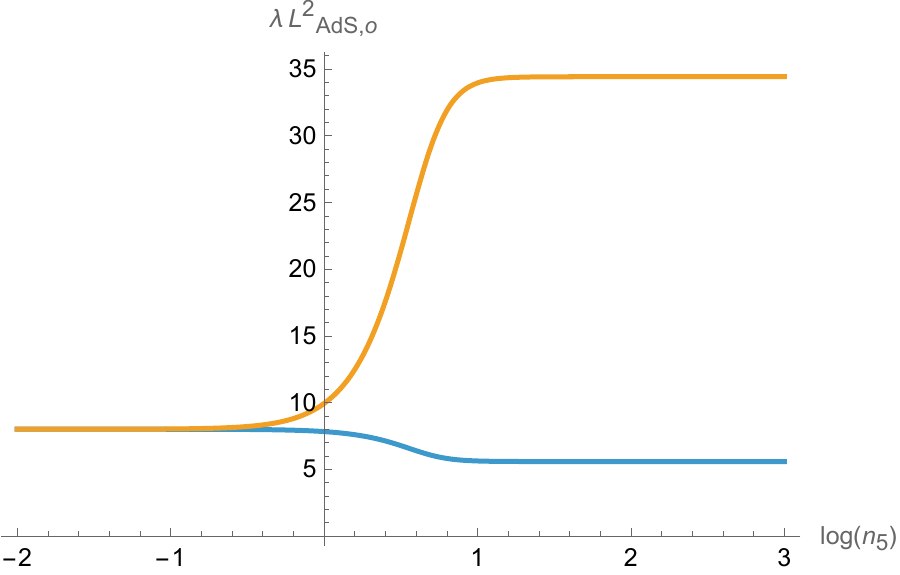}
    \caption{The two eigenvalues for the $\ell=0$ mass matrix plotted against $\log n_{5}$ for $n_{1}=10$.  The tree-level value $L_{AdS,o}^2m^2=8$ is reached at small $n_5$ (equivalently large $n_1$), while for larger values of the ratio $n_5^2/n_1$ the eigenvalues split, but remain positive through the whole range of flux values.}
    \label{fig:l=0 eigenvalues fixed n1}
\end{figure}

\item $(\ell,0)$ tensor: The remaining fields are the traceless spin-two fields.  For $\ell\ge 1$, we can define
\be
X_{\m\n}^{(\ell,0)}=H_{\m\n}^{(\ell,0)}+\frac{1}{L_\circ^{-2}\lp\ell+1\rp^2-\frac{1}{4}\La}\nabla_{\{\m}\nabla_{\n\}}\lp\frac{3}{2}M^{(\ell,0)}-\frac{3}{2}N^{(\ell,0)}+\mathcal{V}^{(\ell,0)}-u^{(\ell,0)}\rp.
\ee
Then using the equations of motion we can verify that $\nabla^\n X_{\m\n}^{(\ell,0)}=0$.  Moreover, substituting into the traceless $g_{\m\n}$ equation and using the equations of motion for the scalar fields, we find that the pieces involving the scalars drop out and we are left with an equation of motion for $X_{\m\n}^{(\ell,0)}$,
\be
\lp\square_0-\lp L_\circ^{-2}\lp\ell^2+2\ell-2\rp+\hlf\La\rp\rp X_{\m\n}^{(\ell,0)}=0,
\ee
corresponding to a massive spin-two field.

For $\ell=0$ we will argue that the residual gauge freedom can be used to entirely gauge away the tensor degrees of freedom.  Indeed, if we first define
\be
X_{\m\n}^{(0,0)}=H_{\m\n}^{(0,0)}+\frac{1}{16L_\circ^{-4}+4L_\circ^{-2}\Lambda-3\Lambda^2}\nabla_{\{\m}\nabla_{\n\}}\lp -\lp 24L_\circ^{-2}+15\La\rp N^{(0,0)}+\lp 32L_\circ^{-2}+18\La\rp\phi^{(0,0)}\rp.
\ee
We also still have some residual gauge transformations parameterized by $\xi_\m^{(0,0)}$ satisfying $\nabla^\m\xi_\m^{(0,0)}=0$.  We can use some of this freedom to impose a gauge condition $\nabla^\n X_{\m\n}^{(0,0)}=0$.  This still leaves some residual freedom, namely parameters $\xi_\m^{(0,0)}$ satisfying
\be
0=\nabla^\n\lp\nabla_\m\xi_\n^{(0,0)}+\nabla_\n\xi_\m^{(0,0)}\rp=\lp\square_0-2L_{AdS,\circ}^{-2}\rp\xi_\m.
\ee
For such $\xi_\m^{(0,0)}$, it is possible to compute that
\be
\lp\square_0+2L_{AdS,\circ}^{-2}\rp\lp\nabla_\m\xi_\n^{(0,0)}+\nabla_\n\xi_\m^{(0,0)}\rp=0.
\ee
With this definition and gauge condition for $X_{\m\n}^{(0,0)}$, the traceless $g_{\m\n}$ equation becomes simply
\be
0=-\lp\square_0+2L_{AdS,\circ}^{-2}\rp X_{\m\n}^{(0,0)}.
\ee
Since this is the same form as the condition on the remaining $\xi_\m^{(0,0)}$ gauge freedom, we can precisely use this freedom to set $X_{\m\n}^{(0,0)}=0$.
\end{itemize}

\subsection{Large $s$ behavior}
\label{subsec:Smalln1}

Using \eqref{small n1 expansions}, can also calculate the eigenvalues of the mass matrix in the large $s$, equivalently small $n_{1}$, limit. For $\ell\geq 1$, these eigenvalues are
\be\label{small n1 eig 1 gen l}
\lim_{n_{1}\rr0}L^{2}_{AdS,o}\la_{1}=2\ell(\ell+2),
\ee
\be\label{small n1 eig 2 gen l}
\lim_{n_{1}\rr0}L^{2}_{AdS,o}\la_{2}=\frac{2}{3}\lp 20+6\ell+3\ell^{2}-2(\xi+\sqrt{3(1-\xi^{2})})\sqrt{2(32+18\ell+9\ell^{2})}\rp,
\ee
\be\label{small n1 eig 3 gen l}
\lim_{n_{1}\rr0}L^{2}_{AdS,o}\la_{3}=\frac{2}{3}\lp 20+6\ell+3\ell^{2}-2(\xi-\sqrt{3(1-\xi^{2})})\sqrt{2(32+18\ell+9\ell^{2})}\rp,
\ee
\be\label{small n1 eig 4 gen l}
\lim_{n_{1}\rr0}L^{2}_{AdS,o}\la_{4}=\frac{2}{3}\lp 20+6\ell+3\ell^{2}+4\xi\sqrt{2(32+18\ell+9\ell^{2})}\rp,
\ee
where
\be
\xi=\frac{1+(\sqrt{2A}\;(115+54\ell+27\ell^{2})+\sqrt{B})^{2/3}}{2(\sqrt{2A}\;(115+54\ell+27\ell^{2})+\sqrt{B})^{1/3}},
\ee
with
\be
A=\frac{1}{(32+18\ell+9\ell^{2})^{3}},\efill B=-\frac{81(78+376\ell+500\ell^{2}+384\ell^{3}+186\ell^{4}+54\ell^{5}+9\ell^{6})}{(32+18\ell+9\ell^{2})^{3}}.
\ee
Although $\sqrt{B}$ is imaginary, $\xi$ can be shown to be real and $\frac{\sqrt{3}}{2}\le\xi\le 1$, so the eigenvalues are all real.  Note that these limiting values are independent of $\la$. For $\ell=2$, we find
\be\label{small n1 eig 1 l=2}
\lim_{n_{1}\rr0}L^{2}_{AdS,o}\la_{1}^{(l=2)}=16,
\ee
\be\label{small n1 eig 2 l=2}
\lim_{n_{1}\rr0}L^{2}_{AdS,o}\la_{2}^{(l=2)}=-0.669,
\ee
\be\label{small n1 eig 3 l=2}
\lim_{n_{1}\rr0}L^{2}_{AdS,o}\la_{3}^{(l=2)}=23.496,
\ee
\be\label{small n1 eig 4 l=2}
\lim_{n_{1}\rr0}L^{2}_{AdS,o}\la_{4}^{(l=2)}=65.173.
\ee
For $\ell=1$, we get (after discarding the unphysical eigenvalue, which is $\la^{(l=1)}_{2}$)
\be\label{small n1 eig 1 l=1}
\lim_{n_{1}\rr0}L^{2}_{AdS,o}\la_{1}^{(l=1)}=6,
\ee
\be\label{small n1 eig 3 l=1}
\lim_{n_{1}\rr0}L^{2}_{AdS,o}\la_{3}^{(l=1)}=13.03,
\ee
\be\label{small n1 eig 4 l=1}
\lim_{n_{1}\rr0}L^{2}_{AdS,o}\la_{4}^{(l=1)}=46.97.
\ee
For $\ell=0$, we get (after discarding the unphysical eigenvalues, which are $\la^{(l=0)}_{1}$ and $\la^{(l=0)}_{2}$)
\be\label{small n1 eig 1 l=0}
\lim_{n_{1}\rr0}L^{2}_{AdS,o}\la_{3}^{(l=0)}=34.4,
\ee
\be\label{small n1 eig 2 l=1}
\lim_{n_{1}\rr0}L^{2}_{AdS,o}\la_{4}^{(l=0)}=5.58,
\ee
These results do match the numerical results shown in figures \ref{fig:l=2 eigenvalues fixed n1}, \ref{fig:l=1 eigenvalues fixed n1}, and \ref{fig:l=0 eigenvalues fixed n1}.
\section{Discussion and outlook}    
In this paper, it is shown that the non-supersymmetric heterotic $O(16)\times O(16)$ compactified on $AdS_{3}\times S^{3}\times T^{4}$ does not uplift to de Sitter due to the one-loop correction to the potential and a generalized argument along the lines of \cite{Basile:2020mpt} is given to show that the absence of this uplift isn't surprising. It is also shown that the scalars and tensors resulting from the deformations of the fields in the six-dimensional effective theory are above the BF bound. These results are quite similar to what was found in~\cite{Baykara:2022cwj}.

There are certain questions that this paper leaves unanswered. One question is regarding the stability under non-perturbative bubbles that change the flux~\cite{Coleman:1977py, Callan:1977pt, Coleman:1980aw,Brown:1988kg}. Such studies have been performed for non-supersymmetric string theories on backgrounds with a single flux~\cite{Antonelli:2019nar}~\footnote{For a nice review, see chapters 5 and 6 of~\cite{Basile:2020xwi}}. In the case of a single flux, the fastest decay channel is when a single flux unit at a time is removed and thus the thin-wall approximation~\cite{Coleman:1977py} would work. It is also known that the `bubble' can be treated as a probe brane (see e.g.~\cite{Antonelli:2019nar}). Multiple flux cases have been discussed in the literature and it has been argued that the fastest decay in this case may be a giant leap instead of a small step and proceeds by a bubble formed from a stack of probe branes instead of a single probe brane~\cite{Bousso:2000xa, Brown:2010mg}. It was also shown in~\cite{Brown:2010mg} that when one has a single flux wrapping on multiple cycles, there are some additional complications that can make those decays dominant that were subdominant in the genuine multiple fluxes case. It was also argued that this case, the decay proceeds by a `minimal' brane that isn't visible in the effective theory.

In the background studied in~\cite{Baykara:2022cwj}, one has magnetic $H_{3}$ flux threading two $S^{3}$ cycles and an electric $H_{3}$ flux on $AdS_{3}$. The background studied in this manuscript one has an electric flux threading on $AdS_{3}$ and a magnetic flux threading an $S^{3}$ cycle. For the case of $\O(1)$ number of fluxes, the issue of whether the fastest decay is a small step or a giant leap would have to be studied case by case and we leave it for future work.

Another direction that we will are exploring in a subsequent article to appear is a more detailed analysis of some specific critical points in the $T^4$ moduli space.  We do not attempt a classification, but there are a set of critical points (determined by rank four even lattices, such as can be obtained by combinations of ADE lattices) where we can do concrete calculations of the mass matrix.  In particular, we will address the open problem of whether any of these critical points can host perturbatively stable examples of the intrinsically quantum vacua (where the electric flux $n_1$ is set to zero), or whether they all develop instabilities in this limit due to some modulus field dropping below the Breitnlohner-Freedman bound.

It would also be of great interest to study these solutions, or the $AdS_3\times S^3\times S^3\times S^1$ solutions of~\cite{Baykara:2022cwj}, purely from the world-sheet perspective, using WZW models for the $AdS_3$ and $S^3$ factors.  Many of the phenomena we uncovered here can also be studied there, without relying on any approximations that the torus direction is at a smaller scale than the spacetime directions.

Finally, it remains true that despite a decent amount of work done \cite{Dixon:1986iz, Ginsparg:1986wr, Alvarez-Gaume:1986ghj, Sugimoto:1999tx, Sagnotti:1995ga, Sagnotti:1996qj, Basile:2021vxh, Mourad:2017rrl, Leone:2025mwo, Dudas:2025ubq, Abel:2015oxa, Blaszczyk:2015zta, Blaszczyk:2014qoa,Mourad:2016xbk, Basile:2018irz, Mourad:2021roa, Mourad:2024dur, Raucci:2025bev, Baykara:2022cwj, Fraiman:2023cpa} to understand the tachyon-free non-supersymmetric string theories, both the $O(16)\times O(16)$ heterotic string, as well as the non-supersymmetric open string models, remain under-studied, and in particular there should be more attention focused on compactifications of these theories. The absence of space-time supersymmetry certainly makes such studies more challenging, but understanding these theories and how they behave in various regimes could potentially help us understand how we might construct stringy solutions that look like our universe.

\section*{Acknowledgments}
The authors would like to thank Ivano Basile, Kaan Baykara, Oleg Lunin, Martin Ro\v{c}ek, and Savdeep Sethi for comments and feedback.  The authors would also like to gratefully acknowledge support from the Simons Center for Geometry and Physics, Stony Brook University for hospitality while part of this research was performed.

\appendix

\section{Theta functions}\label{Appendix A}
The $\Gamma^{20,4}_{\pm}$ lattices used in \eqref{partition function torus 4} are the $d=4$ case for the following lattices;
$$
\Gamma^{d+16,d}_{+}=\{ p:p\in \Gamma^{d+16,d},\;\;p\cdot\d\in \Z\},\efill \Gamma^{d+16,d}_{-}=\{ p:p\in \Gamma^{d+16,d},\;\;p\cdot\d\in \Z/2\}
$$
where $\Gamma^{d+16,d}$ is the usual Narain lattice for $d$ dimensional compactifications for  $E_{8}\times E_{8}$ heterotic theory. We will write down the members of $\Gamma^{d+16,d}$ in the $(p_{L};p_{R};p_{I})$ convention where these three momenta correspond to left momentum, right momentum and the gauge momentum. Frequently, we will refer to $(p_{L};p_{I})$ as $P_{L}$. The vector $\d$ is
$$
\d=\lp 0^{d};0^{d};0^{7},1;0^{7},1\rp.
$$
The theta functions $V_{8},S_{8},O_{8}$ and $C_{8}$ are defined as
$$
V_8=\frac{1}{2\eta^{4}}\lp \vartheta^{4}\ls\begin{matrix}0\\0\end{matrix}\rs(0,\tau)-\vartheta^{4}\ls\begin{matrix}0\\1/2\end{matrix}\rs(0,\tau)
\rp,
$$
$$
S_{8}=\frac{1}{2\eta^{4}}\lp \vartheta^{4}\ls\begin{matrix}1/2\\0\end{matrix}\rs(0,\tau)+\vartheta^{4}\ls\begin{matrix}1/2\\1/2\end{matrix}\rs(0,\tau)
\rp,
$$
$$
O_{8}=\frac{1}{2\eta^{4}}\lp \vartheta^{4}\ls\begin{matrix}0\\0\end{matrix}\rs(0,\tau)+\vartheta^{4}\ls\begin{matrix}0\\1/2\end{matrix}\rs(0,\tau)
\rp,
$$
$$
C_{8}=\frac{1}{2\eta^{4}}\lp \vartheta^{4}\ls\begin{matrix}1/2\\0\end{matrix}\rs(0,\tau)-\vartheta^{4}\ls\begin{matrix}1/2\\1/2\end{matrix}\rs(0,\tau)
\rp,
$$
where $\eta$ is the Dedekind eta function
$$
\eta(\tau)=q^{1/24}\prod_{k=1}^{\infty}\lp 1-q^{k}\rp\;\;\;\text{where }\; q=e^{2\pi i \tau},
$$
and
$$
\vartheta\ls\begin{matrix}
    \al\\\beta\end{matrix}\rs(z,\tau)=\sum_{n\in\Z}\exp\lp \pi i (n+\al)^{2}\tau+2\pi i (n+\al)(n+\beta)\rp.
$$

\section{Gauge lattice sums}\label{Appendix B}
The $E_8$ lattice $\Gamma_{E_8}$ can be described as the union of two sets in $\R^8$,
\begin{equation}
    \Gamma_{E_8}=\Gamma_A\bigcup\Gamma_B,\qquad\Gamma_A=\left\{(n_1,\cdots,n_8)\mid n_i\in\Z,\ \sum_in_i\in2\Z\right\},\quad\Gamma_B=\left\{(r_1,\cdots,r_8)\mid r_i\in\Z+\hlf,\ \sum_ir_i\in 2\Z\right\}.
\end{equation}
The generating function for lengths squared in this lattice can be written down.  The sum over $\Z^8$ is given by
\begin{equation}
    \sum_{v\in\Z^8}q^{\hlf|v|^2}=\lp\sum_{n\in\Z}q^{\hlf n^2}\rp^8=:\vartheta_3(\tau)^8.
\end{equation}
To get the constraint that $\sum_in_i\in 2\Z$ we introduce an analogous sum weighted by the parity of each component,
\begin{equation}
    \sum_{v\in\Z^8}(-1)^{\sum_in_i}q^{\hlf|v|^2}=\lp\sum_{n\in\Z}(-1)^nq^{\hlf n^2}\rp^8=:\vartheta_4(\tau)^8.
\end{equation}
Then we have
\begin{equation}
    \sum_{v\in\Gamma_A}q^{\hlf|v|^2}=\hlf\lp\vartheta_3(\tau)^8+\vartheta_4(\tau)^8\rp.
\end{equation}
Similar analysis gives the generating funxtion for $\Gamma_B$ as
\begin{equation}
    \sum_{v\in\Gamma_B}q^{\hlf|v|^2}=\hlf\lp\sum_{n\in\Z}q^{\hlf\lp n+\hlf\rp^2}\rp^8=\hlf\vartheta_2(\tau)^8.
\end{equation}
Note that the condition $\sum_ir_i\in 2\Z$ in this case is simply taken care of by the factor of $\hlf$, since 
$$
\sum_{n\in\Z}(-1)^nq^{\hlf(n+\hlf)^2}=:\vartheta_1(\tau)=0
$$
Putting these together, we get 
\begin{equation}
    \sum_{v\in\Gamma_{E_8}}q^{\hlf|v|^2}=\hlf\lp\vartheta_3(\tau)^8+\vartheta_4(\tau)^8+\vartheta_2(\tau)^8\rp
\end{equation}

The sums we're actually interested in involve the lattice $\Gamma_{E_8}\oplus\Gamma_{E_8}\subset\R^{16}$ and the additional shift vector $\d=(0,\cdots,0,1;0,\cdots,0,1)\in\R^{16}$.  First we note that the generating function for two copies of the $E_8$ lattice will just be the square of the generating function above.  Next we want the sum weighted by phases $e^{2\pi i\delta\cdot v}$.  If we write $v=v_1\oplus v_2$, then this phase is $+1$ if $v_1$ and $v_2$ are either both in $\Gamma_A$ or both in $\Gamma_B$, or $-1$ if there's one vector in each sublattice.  At the level of generating function, this is accomplished by flipping the signs of the $\vartheta_2(\tau)^8$ terms.  Finally, we want the sum over the lattice shifted by $\d$, weighted by the same phases or not.  The shifting by $\delta$ means that each component $v_i$ is contained in either $\Gamma_A'=\{(n_1,\cdots,n_8)\mid n_i\in\Z,\ \sum n_i\in 2\Z+1\}$ or $\Gamma_B'=\{(r_1,\cdots,r_8)|\mid r_i\in\Z+\hlf,\sum r_i\in 2\Z+1\}$.  This is accomplished by simply flipping the sign of the $\vartheta_4(\tau)^8$ terms in the generating functions.

Summarizing,
\begin{align}
    \sum_{v\in\Gamma_{E_8}\oplus\Gamma_{E_8}}q^{\hlf|v|^2}=\ & \frac{1}{4}\lp\vartheta_3(\tau)^8+\vartheta_4(\tau)^8+\vartheta_2(\tau)^8\rp^2\notag\\
    &\cong 1+480\,q+61920\,q^2+1050240\,q^3+7926240\,q^4+\cdots\\
    \sum_{v\in\Gamma_{E_8}\oplus\Gamma_{E_8}}e^{2\pi i\delta\cdot v}q^{\hlf|v|^2}=\ & \frac{1}{4}\lp\vartheta_3(\tau)^8+\vartheta_4(\tau)^8-\vartheta_2(\tau)^8\rp^2=\vartheta^{8}_{3}(\tau)\vartheta^{8}_{4}(\tau)\notag\\
    &\cong 1-32\,q+480\,q^2-4480\,q^3+29152\,q^4+\cdots\\
    \sum_{v\in\Gamma_{E_8}\oplus\Gamma_{E_8}}q^{\hlf|v+\delta|^2}=\ & \frac{1}{4}\lp\vartheta_3(\tau)^8-\vartheta_4(\tau)^8+\vartheta_2(\tau)^8\rp^2=\vartheta^{8}_{3}(\tau)\vartheta^{8}_{2}(\tau)\notag\\
    &\cong 256\,q+4096\,q^{3/2}+30720\,q^2+147456\,q^{5/2}+\cdots\\
    \sum_{v\in\Gamma_{E_8}\oplus\Gamma_{E_8}}e^{2\pi i\delta\cdot v}q^{\hlf|v+\delta|^2}=\ & \frac{1}{4}\lp\vartheta_3(\tau)^8-\vartheta_4(\tau)^8-\vartheta_2(\tau)^8\rp^2=\vartheta^{8}_{4}(\tau)\vartheta^{8}_{2}(\tau)\notag\\
    &\cong256\,q-4096\,q^{3/2}+30720\,q^2-147456\,q^{5/2} +\cdots
\end{align}
where the representation in terms of the product of theta functions in the last three equations is obtained by using the following quartic identity \cite{mumford2007tata}
\be
\vartheta^{4}_{3}(\tau)-\vartheta^{4}_{2}(\tau)-\vartheta^{4}_{4}(\tau)=0
\ee
\section{$\la$ for square torus}\label{Appendix C}
The partition function for a compactification of $O(16)\times O(16)$ heterotic theory on $T_{4}$ is given as follows;
\be\label{partition function torus 4}
Z_{T_{4}}(\tau)=\frac{1}{\tau^{2}_{2}|\eta|^{8}}\ls \overline{V}_{8}Z_{\G^{20,4}_{+}}-\overline{S}_{8}Z_{\G^{20,4}_{-}}+\overline{O}_{8}Z_{\G^{20,4}_{-}+\d}-\overline{C}_{8}Z_{\G^{20,4}_{+}+\d}\rs
\ee
The lattices $\Gamma^{20,4}_{\pm}$ and the functions $V_{8},S_{8},O_{8}$ and $C_{8}$ are defined in appendix \ref{Appendix A}. The partition functions used in \eqref{partition function torus 4} are defined and simplified as follows;
$$
Z_{\Gamma^{20,4}_{\pm}}=\frac{1}{\eta^{20}\overline{\eta}^{4}}\sum_{p\in\Gamma^{20,4}_{\pm}}q^{\hlf P^{2}_{L}}\overline{q}^{\hlf p^{2}_{R}}=\frac{1}{\eta^{20}\overline{\eta}^{4}}\sum_{p\in\Gamma^{20,4}}\frac{1\pm e^{2\pi i \d\cdot\pi}}{2}q^{\hlf P^{2}_{L}}\overline{q}^{\hlf p^{2}_{R}}
$$
$$
Z_{\Gamma^{20,4}_{\pm}+\d}=\frac{1}{\eta^{20}\overline{\eta}^{4}}\sum_{p\in\Gamma^{20,4}_{\pm}+\d}q^{\hlf P^{2}_{L}}\overline{q}^{\hlf p^{2}_{R}}=\frac{1}{\eta^{20}\overline{\eta}^{4}}\sum_{p\in\Gamma^{20,4}}\frac{1\pm e^{2\pi i \d\cdot\pi}}{2}q^{\hlf \tilde{P}_{L}^{2}}\overline{q}^{\hlf \tilde{p}^{2}_{R}}
$$
where the momenta $\tilde{P}_{L}$ and $\tilde{p}_{R}$ are obtained from $P_{L}$ and $p_{R}$ respectively by sending $\pi$ to $\pi+\d$. The part in the square brackets of \eqref{partition function torus 4} depends on the lattice, and it is given as follows
$$
\frac{1}{\eta^{20}\overline{\eta}^{4}}\lp
\frac{\overline{V}_{8}-\overline{S}_{8}}{2}\underbrace{\sum_{p\in\G^{20,4}}q^{\hlf P^{2}_{L}}\overline{q}^{\hlf p^{2}_{R}}}_{S^{1}_{1}}+\frac{\overline{V}_{8}+\overline{S}_{8}}{2}\underbrace{\sum_{p\in\G^{20,4}}e^{2\pi i \d\cdot\pi}q^{\hlf P^{2}_{L}}\overline{q}^{\hlf p^{2}_{R}}}_{S^{g}_{1}}+\frac{\overline{O}_{8}-\overline{C}_{8}}{2}\underbrace{\sum_{p\in\G^{20,4}}q^{\hlf \tilde{P}_{L}^{2}}\overline{q}^{\hlf \tilde{p}^{2}_{R}}}_{S^{1}_{g}}\right.
$$
\be\label{partition function lattice part}
\left.-\frac{\overline{O}_{8}+\overline{C}_{8}}{2}\underbrace{\sum_{p\in\G^{20,4}}e^{2\pi i \d\cdot\pi}q^{\hlf \tilde{P}_{L}^{2}}\overline{q}^{\hlf \tilde{p}^{2}_{R}}}_{S^{g}_{g}}
\rp
\ee
where we have named different sums to keep track of them. These names are inspired by the terms in the partition function of the $S^{1}/\Z_{2}$ CFT. The first term in this bracket vanishes because $V_{8}-S_{8}=0$. Therefore, the expression for the torus partition function at the extremum point is as follows
\be
Z_{T_{4}}(\tau)=\frac{1}{2\tau^{2}_{2}\eta^{24}\overline{\eta}^{8}}\ls
\lp\overline{V}_{8}+\overline{S}_{8}\rp S^{g}_{1}+\lp\overline{O}_{8}-\overline{C}_{8}\rp S^{1}_{g}-\lp\overline{O}_{8}+\overline{C}_{8}\rp S^{g}_{g}\rs
\ee
Using \eqref{lambda general expression} for the square torus, we have the following;
\be\label{lambda expression}
\la=-\frac{\pi}{2(2\pi)^{4}}\int\frac{d^{2}\tau}{4\tau^{4}_{2}\eta^{24}\overline{\eta}^{12}}\ls
\lp\tilde{\overline{V}}_{8}+\tilde{\overline{S}}_{8}\rp S^{g}_{1}+\lp\tilde{\overline{O}}_{8}-\tilde{\overline{C}}_{8}\rp S^{1}_{g}-\lp\tilde{\overline{O}}_{8}+\tilde{\overline{C}}_{8}\rp S^{g}_{g}\rs
\ee
where the tilde on theta functions indicates that the factor of $(2\overline{\eta}^{4})^{-1}$ that appears in their definitions have been taken out. To use this formula, we will need some series expansions. The series for the $(\eta^{24}\overline{\eta}^{12})^{-1}$ factor is as follows
$$
\frac{1}{\eta^{24}\overline{\eta}^{12}}=e^{-\pi i\tau_{1}}e^{3\pi \tau_{2}}+12(2e^{\pi i\tau_{1}}+e^{-3\pi i\tau_{1}})e^{\pi \tau_{2}}+18(5 e^{-5\pi i\tau_{1}}+16 e^{-\pi i\tau_{1}}+18 e^{3\pi i\tau_{1}})e^{-\pi\tau_{2}}
$$
\be\label{eta combination expansion}
+8(65 e^{-7\pi i\tau_{1}}+270 e^{-3\pi i\tau_{1}}+486e^{\pi i\tau_{1}}+400 e^{5\pi i\tau_{1}})e^{-3\pi\tau_{2}}+...
\ee
which means that if we want the integrand in \eqref{lambda expression} up to the $e^{-\pi\tau_{2}}$ term, we need the bracketed expression in \eqref{lambda expression} up to the $e^{-4\pi\tau_{2}}$ term. We now turn to the theta functions. They are expanded as follows
\be
\tilde{\overline{V}}_{8}+\tilde{\overline{S}}_{8}=32e^{-\pi i \tau_{1}}e^{-\pi \tau_{2}}+128 e^{-3\pi i \tau_{1}}e^{-3\pi \tau_{2}}+\O\lp e^{-5\pi \tau_{2}}\rp
\ee
\be
\tilde{\overline{O}}_{8}\pm \tilde{\overline{C}}_{8}=2\pm 16 e^{-\pi i \tau_{1}}e^{-\pi \tau_{2}}+48 e^{-2\pi i \tau_{1}}e^{-2\pi \tau_{2}}\pm 64 e^{-3\pi i \tau_{1}}e^{-3\pi \tau_{2}}+48 e^{-4\pi i \tau_{1}}e^{-4\pi \tau_{2}}+\O(e^{-5\pi \tau_{2}})
\ee
Lastly, let's work out the expansions of the sums $S^{1}_{1},S^{g}_{1},S^{1}_{g}$ and $S^{g}_{g}$ which will require more work. These sums are products of a gauge lattice part (which isn't the same for all of these sums) and a momentum/winding part which is common. This common part is given as follows
\be\label{nw sum for square torus}
\lp\sum_{n,w}e^{-\pi \tau_{2}(n^{2}+w^{2})}e^{2\pi i\tau_{1}nw}\rp^{4}=\ls 1+4e^{-\pi\tau_{2}}+4\cos(2\pi \tau_{1}) e^{-2\pi\tau_{2}}+4e^{-4\pi\tau_{2}}+\O(e^{-5\pi\tau_{2}})\rs^{4}
\ee
where we only needed to sum over the following $(n,w)$ pairs to get the $e^{-4\pi\tau_{2}}$ terms
$$
(n,w)=(0,0),(0,\pm1),(\pm1,0),(\pm1, \pm 1), (\pm 1,\mp 1),(\pm 2,0),(0,\pm 2).
$$
The series expansions of the gauge lattice part of $S^{1}_{1},S^{g}_{1},S^{1}_{g}$ and $S^{g}_{g}$ are calculated by the analysis in Appendix \ref{Appendix B} and they are given as follows
\begin{align}
&S^{1}_{1}:\;\;\sum_{\pi}e^{\pi i \tau_{1}|\pi|^{2}}e^{-\pi\tau_{2}|\pi|^{2}}=1+480e^{2\pi i \tau_{1}}e^{-2\pi\tau_{2}}+61920e^{4\pi i \tau_{1}}e^{-4\pi\tau_{2}}+...\\
&S^{g}_{1}:\;\;\sum_{\pi}e^{2\pi i \d.\pi}e^{\pi i \tau_{1}|\pi|^{2}}e^{-\pi\tau_{2}|\pi|^{2}}=1-32e^{2\pi i \tau_{1}}e^{-2\pi\tau_{2}}+480e^{4\pi i \tau_{1}}e^{-4\pi\tau_{2}}+...\\
&S^{1}_{g}:\;\;\sum_{\pi}e^{\pi i \tau_{1}|\pi+\d|^{2}}e^{-\pi\tau_{2}|\pi+\d|^{2}}=256e^{2\pi i \tau_{1}}e^{-2\pi\tau_{2}}+4096e^{3\pi i \tau_{1}}e^{-3\pi\tau_{2}}+30720e^{4\pi i \tau_{1}}e^{-4\pi\tau_{2}}
+...\\
&S^{g}_{g}:\;\;\sum_{\pi}e^{2\pi i\d.\pi}e^{\pi i \tau_{1}|\pi+\d|^{2}}e^{-\pi\tau_{2}|\pi+\d|^{2}}=256e^{2\pi i \tau_{1}}e^{-2\pi\tau_{2}}-4096e^{3\pi i \tau_{1}}e^{-3\pi\tau_{2}}+30720e^{4\pi i \tau_{1}}e^{-4\pi\tau_{2}}+...
\end{align}
Now, we can expand the integrand in \eqref{lambda expression} up to $\O(e^{-\pi\tau_{2}})$ and calculate the value of $\la$ for the square torus with vanishing $b$ field and Wilson lines. We get the following;
$$
\la=-\frac{\pi}{2(2\pi)^{4}}\int_{\mathcal{F}}\frac{d^{2}\tau_{2}}{4\tau^{4}_{2}}\ls 32e^{-2\pi i \tau_{1}}e^{2\pi \tau_{2}}+512e^{-2\pi i \tau_{1}}e^{\pi\tau_{2}}+256\lp 3 e^{-4\pi i \tau_{1}}+12 e^{-2\pi i \tau_{1}}-32\rp\right.
$$
\be\label{lambda value}
\left.+1024\lp 11 e^{-4\pi i \tau_{1}}+8 e^{-2\pi i \tau_{1}}+16e^{3\pi i \tau_{1}}-129\rp e^{-\pi\tau_{2}}+...\rs\sim 1.321
\ee

\section{First derivatives}\label{Appendix D}

We need to calculate the first derivatives of \eqref{partition function torus 4} with respect to the metric, $b$ field (where $b$ is the background Kalb Ramond field) and the Wilson line moduli, and show that they are zero for square torus, with vanishing $b$ field and Wilson lines. The expressions for the square of momenta are as follows \cite{Narain:1986am};
\be\label{PL^2}
P^{2}_{L}=p^{2}_{L}+p^{2}_{I}=\hlf g^{ik}\lp n_{i}+\hat{E}_{ij}w^{j}-\pi\cdot A_{i}\rp\lp n_{k}+\hat{E}_{kl}w^{l}-\pi\cdot A_{k}\rp+\d_{IJ}\lp \pi^{I}+A^{I}_{i}w^{i}\rp\lp \pi^{J}+A^{J}_{j}w^{j}\rp
\ee
\be\label{pR^2}
p^{2}_{R}=\hlf g^{ik}\lp n_{i}-E_{ij}w^{j}-\pi\cdot A_{i}\rp\lp n_{k}-E_{kl}w^{l}-\pi\cdot A_{k}\rp
\ee
where we have;
\be
\hat{E}_{ij}=g_{ij}+b_{ij}+\hlf A_{i}\cdot A_{j}\efill\hat{E}_{ij}=2g_{ij}-E_{ij}=g_{ij}-b_{ij}-\hlf A_{i}\cdot A_{j}
\ee
The expressions for $\tilde{P}^{2}_{L}$ and $\tilde{p}^{2}_{R}$ are quite similar;
\be\label{tildePL2 pR2}
\tilde{P}^{2}_{L}=P^{2}_{L}\text{ with }\pi\rr\pi+\d\efill\efill\tilde{p}^{2}_{R}=p^{2}_{R}\text{ with }\pi\rr\pi+\d
\ee
The derivatives of the squared momenta in \eqref{PL^2} and \eqref{pR^2} are as follows
\be
\frac{\p p^{2}_{R}}{\p A^{I}_{q}}=-\lp \pi_{I}g^{qk}+\hlf (g^{qk}A_{Ij}w^{j}+g^{ik}w^{q}A_{Ii})\rp(n_{k}-E_{kl}w^{l}-\pi\cdot A_{k})
\ee
\be
\frac{\p P^{2}_{L}}{\p A^{I}_{q}}=2 w^{q}(\pi_{I}+A_{Ij}w^{j})-\lp \pi_{I}g^{qk}+\hlf (g^{qk}A_{Ij}w^{j}+g^{ik}w^{q}A_{Ii})\rp(n_{k}+\hat{E}_{kl}w^{l}-\pi\cdot A_{k})
\ee
\be
\frac{\p p^{2}_{R}}{\p g_{pq}}=-\frac{1}{2} g^{ip}g^{kq}(n_{i}-E_{ij}w^{j}-\pi.A_{i})(n_{k}-E_{kl}w^{l}-\pi\cdot A_{k})-\hlf \lp
g^{pk}w^{q}+g^{qk}w^{p}\rp (n_{k}-E_{kl}w^{l}-\pi\cdot A_{k})
\ee
\be
\frac{\p P^{2}_{L}}{\p g_{pq}}=-\frac{1}{2} g^{ip}g^{kq}(n_{i}+\hat{E}_{ij}w^{j}-\pi.A_{i})(n_{k}+\hat{E}_{kl}w^{l}-\pi\cdot A_{k})+\hlf \lp
g^{pk}w^{q}+g^{qk}w^{p}\rp (n_{k}+\hat{E}_{kl}w^{l}-\pi\cdot A_{k})
\ee
\be
\frac{\p p^{2}_{R}}{\p b_{pq}}=-\hlf \lp
g^{pk}w^{q}-g^{qk}w^{p}\rp (n_{k}-E_{kl}w^{l}-\pi\cdot A_{k})
\ee
\be
\frac{\p P^{2}_{L}}{\p b_{pq}}=-\hlf \lp
g^{pk}w^{q}-g^{qk}w^{p}\rp (n_{k}+\hat{E}_{kl}w^{l}-\pi\cdot A_{k})
\ee
while the derivatives of $\tilde{P}^{2}_{L}$ and $\tilde{p}^{2}_{R}$ are obtained from $P^{2}_{L}$ and $p^{2}_{R}$ respectively by sending $\pi$ to $\pi+\d$. For the square torus with vanishing $b$ field and Wilson lines, these derivatives become the following
\be
\left.\frac{\p P^{2}_{L}}{\p A^{I}_{q}}\right|=\left.\frac{\p p^{2}_{R}}{\p A^{I}_{q}}\right|=-(n^{q}-w^{q})\pi_{I}\efill\efill \left.\frac{\p \tilde{P}_{L}^{2}}{\p A^{I}_{q}}\right|=-\left.\frac{\p \tilde{p}^{2}_{R}}{\p A^{I}_{q}}\right|=(\pi+\d)_{I}\lp n^{q}-w^{q}\rp
\ee
\be
\left.\frac{\p P^{2}_{L}}{\p g_{pq}}\right|=\left.\frac{\p p^{2}_{R}}{\p g_{pq}}\right|=\left.\frac{\p \tilde{P}^{2}_{L}}{\p g_{pq}}\right|=\left.\frac{\p \tilde{p}^{2}_{R}}{\p g_{pq}}\right|=-\hlf (n^{p}n^{q}-w^{p}w^{q})
\ee
\be
\left.\frac{\p P^{2}_{L}}{\p b_{pq}}\right|=\left.\frac{\p p^{2}_{R}}{\p b_{pq}}\right|=\left.\frac{\p \tilde{P}^{2}_{L}}{\p b_{pq}}\right|=\left.\frac{\p \tilde{p}^{2}_{R}}{\p b_{pq}}\right|=-\hlf\lp n^{p}w^{q}-n^{q}w^{p}\rp
\ee
where the vertical line indicates that we are evaluating the quantity at the specified point in the moduli space. Lastly, we will need the following expressions;
$$
\left.P^{2}_{L}\right|=\hlf \lp n^{2}+2n\cdot w+w^{2}\rp+\d_{IJ}\pi^{I}\pi^{J}\efill \left.p^{2}_{R}\right|=\hlf \lp n^{2}-2n\cdot w+w^{2}\rp
$$
$$
\left.\tilde{P}^{2}_{L}\right|=\hlf \lp n^{2}+2n\cdot w+w^{2}\rp+\d_{IJ}(\pi+\d)^{I}(\pi+\d)^{J}\efill \left.\tilde{P}^{2}_{R}\right|=\hlf \lp n^{2}-2n\cdot w+w^{2}\rp
$$
where we used the following shorthands;
$$
n^{2}=g^{ij}n_{i}n_{j}\efill w^{2}=g_{ij}w^{i}w^{j}\efill n.w=n_{i}w^{i}.
$$
Using the above results, we get the following expressions for the $A^{I}_{i}$ derivatives of the four terms in \eqref{partition function lattice part}
\be
\left.\frac{\p S^{1}_{1}}{\p A^{I}_{i}}\right|=2\pi\tau_{2}\sum_{n,w,\pi}\pi_{I}(n^{i}-w^{i})q^{\hlf \left.P^{2}_{L}\right|}\;\overline{q}^{\hlf \left.p^{2}_{R}\right|}
\ee
\be
\left.\frac{\p S^{g}_{1}}{\p A^{I}_{i}}\right|=2\pi\tau_{2}\sum_{n,w,\pi}e^{2\pi i\d.\pi}\pi_{I}(n^{i}-w^{i})q^{\hlf \left.P^{2}_{L}\right|}\;\overline{q}^{\hlf \left.p^{2}_{R}\right|}
\ee
\be
\left.\frac{\p S^{1}_{g}}{\p A^{I}_{i}}\right|=2\pi\tau_{2}\sum_{n,w,\pi}(\pi+\d)_{I}(n^{i}-w^{i})q^{\hlf \left.\tilde{P}^{2}_{L}\right|}\;\overline{q}^{\hlf \left.\tilde{p}^{2}_{R}\right|}
\ee
\be
\left.\frac{\p S^{g}_{g}}{\p A^{I}_{i}}\right|=2\pi\tau_{2}\sum_{n,w,\pi}e^{2\pi i\d.\pi}(\pi+\d)_{I}(n^{i}-w^{i})q^{\hlf \left.\tilde{P}^{2}_{L}\right|}\;\overline{q}^{\hlf \left.\tilde{p}^{2}_{R}\right|}
\ee
All of these derivatives are zero and it can be seen by renaming $n$'s and $w$'s (by swapping them). Note that we can do this swapping without changing $P^{2}_{L}|,p^{2}_{L}|,\tilde{P}^{2}_{L}|$ or $\tilde{p}^{2}_{R}|$. This wouldn't necessarily be the case if we were at some other point in the moduli space. The metric derivatives are given as follows;
\be
\left.\frac{\p S^{1}_{1}}{\p g_{pq}}\right|=\pi\tau_{2}\sum_{n,w,\pi}(n^{p}n^{q}-w^{p}w^{q})q^{\hlf \left.P^{2}_{L}\right|}\;\overline{q}^{\hlf \left.p^{2}_{R}\right|}
\ee
\be
\left.\frac{\p S^{g}_{1}}{\p g_{pq}}\right|=\pi\tau_{2}\sum_{n,w,\pi}e^{2\pi i\d\cdot\pi}(n^{p}n^{q}-w^{p}w^{q})q^{\hlf \left.P^{2}_{L}\right|}\;\overline{q}^{\hlf \left.p^{2}_{R}\right|}
\ee
\be
\left.\frac{\p S^{1}_{g}}{\p g_{pq}}\right|=\pi\tau_{2}\sum_{n,w,\pi}(n^{p}n^{q}-w^{p}w^{q})q^{\hlf \left.\tilde{P}^{2}_{L}\right|}\;\overline{q}^{\hlf \left.\tilde{p}^{2}_{R}\right|}
\ee
\be
\left.\frac{\p S^{g}_{g}}{\p g_{pq}}\right|=\pi\tau_{2}\sum_{n,w,\pi}e^{2\pi i\d\cdot\pi}(n^{p}n^{q}-w^{p}w^{q})q^{\hlf \left.\tilde{P}^{2}_{L}\right|}\;\overline{q}^{\hlf \left.\tilde{p}^{2}_{R}\right|}
\ee
These derivatives also vanish if we swap $n$'s with $w$'s. Lastly, the $b$ field derivatives are as follows
\be
\left.\frac{\p S^{1}_{1}}{\p b_{pq}}\right|=\pi\tau_{2}\sum_{n,w,\pi}(n^{p}w^{q}-n^{q}w^{p})q^{\hlf \left.P^{2}_{L}\right|}\;\overline{q}^{\hlf \left.p^{2}_{R}\right|}
\ee
\be
\left.\frac{\p S^{g}_{1}}{\p b_{pq}}\right|=\pi\tau_{2}\sum_{n,w,\pi}e^{2\pi i\d.\pi}(n^{p}w^{q}-n^{q}w^{p})q^{\hlf \left.P^{2}_{L}\right|}\;\overline{q}^{\hlf \left.p^{2}_{R}\right|}
\ee
\be
\left.\frac{\p S^{g}_{1}}{\p b_{pq}}\right|=\pi\tau_{2}\sum_{n,w,\pi}(n^{p}w^{q}-n^{q}w^{p})q^{\hlf \left.\tilde{P}^{2}_{L}\right|}\;\overline{q}^{\hlf \left.\tilde{p}^{2}_{R}\right|}
\ee
\be
\left.\frac{\p S^{g}_{g}}{\p b_{pq}}\right|=\pi\tau_{2}\sum_{n,w,\pi}e^{2\pi i\d.\pi}(n^{p}w^{q}-n^{q}w^{p})q^{\hlf \left.\tilde{P}^{2}_{L}\right|}\;\overline{q}^{\hlf \left.\tilde{p}^{2}_{R}\right|}
\ee
and as expected, these derivatives also vanish by using the $n\leftrightarrow w$ swap. Therefore, the square torus with vanishing $b$ field and Wilson lines is an extremum point in the moduli space.

\printbibliography
\end{document}